\documentclass[twocolumn,showpacs,preprintnumbers,amsmath,amssymb]{revtex4-1}
\usepackage{textcomp}
\usepackage[pdftex]{epsfig}
\DeclareGraphicsExtensions{.jpg,.mps,.pdf,.png}

\usepackage{dcolumn}
\usepackage{bm}

\newcommand{\aap}{{\em Astr. \& Astrophys. \/}}

\newcommand{\physrep}{{\em Phys. Rep. \/}}

\newcommand{\vx}{{\bf x}}
\newcommand{\vq}{{\bf q}}
\newcommand{\vv}{{\bf v}}
\newcommand{\vk}{{\bf k}}
\newcommand{\dd}{{\rm d}}
\newcommand{\bb}{\alpha}

\newcommand{\mD}{{\cal D}}
\newcommand{\cdm}{{\rm c}}
\newcommand{\ba}{{\rm b}}

\newcommand{\Dirac}{\delta_{\rm D}}

\newcommand{\rhob}{\overline{\rho}}

\newcommand{\xv}{{\bf x}}

\newcommand{\displ}{{\rm d}}

\newcommand{\ci}{{\rm ci}}
\newcommand{\ii}{{\rm is}}

\newcommand{\init}{{\rm in}}

\newcommand{\lon}{{\rm lon}}
\newcommand{\HH}{{\cal H}}

\newcommand{\rtheta}{{\Theta}}

\newcommand{\egamma}{\gamma}
\newcommand{\tPsi}{\tilde{\Psi}}

\newcommand{\sigmax}{\sigma_{\times}}

\renewcommand{\[}{\begin{equation}}
\renewcommand{\]}{\end{equation}}

\begin{document}

\title{Resummed propagators in multi-component cosmic fluids\\ with the \textsl{eikonal}  approximation}

\author{Francis Bernardeau}
\email{francis.bernardeau@cea.fr}
\author{Nicolas Van de Rijt}
\email{nicolas.van-de-rijt@cea.fr}
\author{Filippo Vernizzi}
\email{filippo.vernizzi@cea.fr}
\affiliation{
Institut de Physique Th{\'e}orique,
CEA, IPhT, F-91191 Gif-sur-Yvette, France
CNRS, URA 2306, F-91191 Gif-sur-Yvette, France
}
\vspace{.2 cm}
\date{\today}
\vspace{.2 cm}
\begin{abstract}
We introduce the \textsl{eikonal} approximation to study the effect of the large-scale motion of cosmic fluids on their small-scale evolution. This approach consists in collecting the impact of the long-wavelength displacement field into a single or finite number of random variables, whose statistical properties can be computed from the initial conditions. For a single dark matter fluid, we show that we can recover the nonlinear propagators of renormalized perturbation theory. These are obtained with no need  to assume that the displacement field follows the linear theory. Then we extend the eikonal approximation to many fluids. In particular, we study the case of two non-relativistic components and we derive their resummed propagators in the presence of isodensity modes. Unlike the adiabatic case, where only the phase of small-scale modes is affected by the large-scale advection field, the isodensity modes change also the amplitude on small scales. We explicitly solve the case of cold dark matter-baryon  mixing and find that the isodensity modes induce only very small  corrections to the resummed propagators. 
\end{abstract}
 \vskip2pc

\maketitle

\section{Introduction}

The development of wide-field surveys has triggered renewed interest in the implementation of perturbation techniques for the computation of the statistical properties of large-scale structures. Several approaches have been proposed to significantly extend the standard perturbation theory (PT) methods (see \cite{2002PhR...367....1B}). A particularly interesting approach is the so-called renormalized perturbation theory (RPT), pioneered by Crocce and Scoccimarro \cite{2006PhRvD..73f3519C,2006PhRvD..73f3520C,2007arXiv0704.2783C}. This method relies on the use of the 2-point propagator as a measure of the memory of the initial conditions. This appears as the fundamental building block from which perturbation theory can be re-constructed and allows to take into account nonlinearities from very small scales, reducing their impact in the neglected terms of the perturbative expansion. This idea was later extended in \cite{2008PhRvD..78j3521B,2010PhRvD..82h3507B} with the introduction of multi-point propagators. 

The key result of RPT is that in the high-$k$ limit the propagators can be computed exactly, by summing up an infinite subset of contributions in the standard perturbation theory expansion. However, this result has been proved only for a single pressureless fluid -- describing cold dark matter (CDM) -- using a technique that seems difficult to extend to more complex scenarios. Thus, there is no systematic way to implement the RPT approach when the content of the cosmic fluid is richer -- e.g.~when it includes various matter components, non-relativistic neutrinos, or even modification of gravity -- and its application range has been so far limited to simple cosmological models (see however \cite{2010PhRvD..81b3524S}). Note that other approaches, such as the so-called time renormalization group approach \cite{2008JCAP...10..036P}, do not suffer from such limitations. 

On the other hand,  the resummation of the propagators can be obtained by a more direct technique than that originally introduced in \cite{2006PhRvD..73f3519C,2006PhRvD..73f3520C,2007arXiv0704.2783C}. As mentioned in \cite{2008A&A...484...79V} and explicitly used in \cite{2008PhRvD..78h3503B,2010PhRvD..82h3507B}, in the high-$k$ limit it is possible to resum the same class of contributions making use of a single or a finite number of random variables, which describe the effect of the long-wavelength fluctuations on smaller scales. This has been called the $\alpha$-method \cite{2010PhRvD..82h3507B}. Here we will explicitly present how to compute nonlinear propagators in this framework. Moreover, we will show that this method can be employed to extend the RPT approach to arbitrarily complicated cosmologies. Borrowing the terminology from quantum field theory, where similar techniques are used (see e.g.~\cite{Abarbanel:1969ek,Levy:1969cr}), we propose to dub this method the \textsl{eikonal approximation}. 

The fluid content of the universe is richer than a simple single-dark matter component. In practice we know that the properties of the large-scale structure of the universe can be significantly affected by the presence of a sub-dominant species. This is the case of baryons, which at high enough redshift behave very differently from CDM. Indeed, the net result of this different behavior is the existence of the baryonic oscillations. 

From a theoretical point of view, the CDM-baryon system is very appealing: After decoupling both components are pressureless fluids (at least above the baryonic Jeans scale) and thus follow geodesic motion \footnote{Another case studied in PT with two components following geodesic motion is a mixture of CDM and clustering quintessence \cite{2011JCAP...03..047S}. However, in this case the initial conditions are such that the two fluids remain comoving during their evolution and isodensity modes do not develop.}. However, the matter fluid as a whole cannot be described as a pressureless effective fluid. The reason is that the CDM and baryon fluids are moving at different velocities -- see \cite{2010PhRvD..82h3520T} for a study of some of the consequences of this different behavior and their possible observational implications. Such a velocity dispersion induces an effective anisotropic pressure in the total fluid, modifying its equations of motion. Thus, one is forced to study the system of coupled equations for the CDM and baryons, as previously done in \cite{2010PhRvD..81b3524S}. In particular, this is the system that we will explore with the help of the eikonal approximation.

The plan of the paper is the following. In Section \ref{Single-fluid} we review the basic concepts of the RPT approach for a single CDM fluid and we discuss the eikonal approximation in this context. In Section \ref{Multi-fluids} we extend this discussion to the multi-fluid case. In particular, we derive the evolution equations for several gravitationally coupled pressureless fluids, we describe the various modes that appear in this case, and we present how they can be incorporated in the eikonal approximation. Finally, in Section \ref{CDMbaryons} we illustrate our concepts in the case of the standard CDM-baryon mixing.

\section{Single fluid}
\label{Single-fluid}

In this section we review the basic concepts of the RPT approach with a single perfect fluid, developed in \cite{2006PhRvD..73f3519C,2006PhRvD..73f3520C}. Moreover, we rederive the procedure to resum the nonlinear propagators for cosmic fluids using the eikonal approximation.

\subsection{Equations of motion}

We assume the universe to be filled by one pressureless fluid. We denote its density by $\rho$,  and the density contrast by $\delta \equiv \rho/\rhob-1$, where $\rhob$ is the average energy density. The continuity equation then reads
\[
\frac{\partial}{\partial t} \delta+\frac1a\left((1+\delta)u^i \right)_{,i}=0 \;,
\label{continuity}
\]
where $u^{i}$ is the $i$-component of the peculiar velocity field of the fluid and a comma denotes the partial derivative. 
The Euler equation is
\[
\frac{\partial}{\partial t}u^{i}+H u^{i}+\frac{1}{a}u^{j} u^i_{,j}=-\frac{1}{a}\phi_{,i} \;,
\label{Euler_tot}
\]
where $H$ is the Hubble rate, $H \equiv \dd \ln a /\dd t$, and $\phi$ is the gravitational potential. Since we are only interested in the dynamics on sub-horizon scales, $\phi$ is the usual Newtonian potential, satisfying the  Poisson equation
\[
\Delta\phi=4\pi G\,a^{2} \rhob \delta \;.
\label{poisson}
\]

We also ignore small-scale shell crossings in the fluid. Then, since the gravitational force is potential the fluid velocity remains potential at all orders in the perturbations. Thus, it can be entirely described by the dimensionless velocity divergence, defined by
\[
\theta \equiv \frac{u^i_{,i}}{aH} \label{def_theta} \;.
\]

By using the following convention for the Fourier modes, 
\[
f(\vk)\equiv\int\frac{\dd^3\vx}{(2\pi)^3}f (\vx)e^{-i \vk \cdot\vx} \;,
\]
the equations of motion can then be rewritten in Fourier space as
\begin{widetext}
\begin{align}
\frac{1}{H}\frac{\partial}{\partial t}\delta (\vk)+\theta(\vk)&
=- \alpha(\vk_{1},\vk_{2})\theta(\vk_{1})\delta(\vk_{2})
\label{Cont1} \;,
\\
\frac{1}{H}\frac{\partial}{\partial t}\theta(\vk)
+\frac{1}{H}\frac{\dd \ln (a^2 H)}{\dd t}\theta(\vk)
+\frac{3}{2}\Omega_{\rm m}\delta (\vk)&
=- \beta(\vk_{1},\vk_{2})\theta(\vk_{1})\theta(\vk_{2})
\label{Eul1} \;,
\end{align}
\end{widetext}
where $\Omega_{\rm m}$ is the reduced matter density and
\begin{align}
\alpha(\vk_{1},\vk_{2}) &=\frac{(\vk_{1}+\vk_{2})\cdot \vk_{1}}{k_{1}^2} \;,
\label{alpha}
\\
\beta(\vk_{1},\vk_{2}) &=\frac{(\vk_{1}+\vk_{2})^2\ \vk_{1}\cdot \vk_{2}}{2k_{1}^2 k_{2}^2}\;.
\label{beta}
\end{align}
On the right-hand side of eqs.~(\ref{Cont1}) and (\ref{Eul1}), integration over repeated wave modes and a Dirac function  $\Dirac(\vk-\vk_{1}-\vk_{2})$ is implied. Note that these equations are valid irrespective of the dark energy equation of state or curvature term.

\subsection{RPT formulation}

In order to recast these equations in RPT form let us first discuss their linear solutions. At linear order, the coupling terms in the right-hand side of eqs.~\eqref{Cont1} and \eqref{Eul1} are absent. We are then left with the usual linear solutions of a pressureless fluid, i.e.
\[
\delta(\vx,t)=D_{+}(t)\delta_{+}(\vx)+D_{-}(t)\delta_{-}(\vx)\;, \label{delta+-}
\]
where $D_{+}(t)$ and $D_{-}(t)$ correspond to the growing and decaying modes, respectively. The corresponding expression for the dimensionless velocity divergence is
\[
\theta(\vx,t)=-f_{+}(t)D_{+}(t)\delta_{+}(\vx)-f_{-}(t)D_{-}(t)\delta_{-} (\vx)\;, \label{theta+-}
\]
where $f_{+}$ and $f_{-}$ are the growth rates, defined as $f_\pm \equiv \dd \ln D_\pm /\dd \ln a$.

As introduced in \cite{2001NYASA.927...13S,2010PhRvD..81b3524S}, it is convenient to define the duplet  
\[
\Psi_{a}=
\left(\begin{array}{c}
\delta   \\  \rtheta \end{array} \right)
 \;, \label{Psi_def}
\]
where $\rtheta$ is the reduced velocity contrast defined as
\[
\begin{split}
\rtheta (\vx,t)&\equiv -{\theta}(\vx,t)/{f_{+}(t)} \\
&= D_{+}(t)\delta_{+}(\vx) + \frac{f_-(t)}{f_+(t)}D_{-} (t) \delta_{-} (\vx) \;, \label{Theta+-} 
\end{split}
\]
in such a way that the linear growing mode of $\Theta$ is the same as that of $\delta$. It is then convenient to rewrite the evolution equation using $\eta$ as time variable, defined through
\[
D_{+}\dd\eta \equiv \dd D_{+}\;.
\]
With this definition the equations of motion \eqref{Cont1} and \eqref{Eul1} can be recast as
\[
\frac{\partial}{\partial \eta}\Psi_{a}(\vk)+\Omega_{ab}\Psi_{b}(\vk)= \gamma_{abc}(\vk,\vk_{1},\vk_{2})\Psi_{b}(\vk_{1})\Psi_{c}(\vk_{2}) \;, \label{EOM}
\]
where 
\[
\Omega_{ab} \equiv 
\left(\begin{array}{cc}
 0 & -1  \\
 -\frac{3}{2}\frac{\Omega_{m}}{f_{+}^{2}} & \frac{3}{2}\frac{\Omega_{m}}{f_{+}^{2}}-1  
 \end{array} \right)\;,
\]
and the non-zero elements of the coupling matrix $\gamma_{abc}$ are
\[
\begin{split}
\gamma_{112}(\vk,\vk_{1},\vk_{2})&\equiv \frac{\alpha(\vk_{2},\vk_{1})}{2}\;,\\
\gamma_{121}(\vk,\vk_{1},\vk_{2})&\equiv \frac{\alpha(\vk_{1},\vk_{2})}{2}\;,\\
\gamma_{222}(\vk,\vk_{1},\vk_{2})&\equiv \beta(\vk_{1},\vk_{2}) \;. \label{gamma}
\end{split}
\]

From eqs.~\eqref{delta+-} and \eqref{Theta+-}, the growing and decaying solutions are proportional to
\[
u_{a}^{(+)}\propto\left(1,1\right)^T\;\quad\mathrm{and}\quad u_{a}^{(-)}\propto(1,{f_{-}}/{f_{+}})^T \label{eq:ic}
\]
respectively. It has been widely stressed that $f_{-}/f_{+}$ is very weakly dependent on the background. Indeed, it departs little from the value it takes in an Einstein-de Sitter universe (EdS), i.e.~$\Omega_m =1$, where $f_{-}/f_{+}=-3/2$ \footnote{It is straightforward to show that for $\Lambda$CDM $f_- = -\frac32 \Omega_m$ and $f_+/f_- = 1 -\frac53 \frac{a}{D_+}$.}.

The solutions of the linear equations of motion, obtained from dropping the right-hand side of \eqref{EOM}, can be formally written in terms of the \textsl{linear} propagator $g_{ab}(\eta,\eta')$.
This is such that
\[
g_{ab}(\eta,\eta)=\delta_{ab} \;,
\]
and
\[
\frac{\partial}{\partial \eta}g_{ab}(\eta,\eta_0)+\Omega_{ac} (\eta) g_{cb}(\eta,\eta_0)=0 \;.
\]
It can be built from a complete set of independent solutions of the evolution equation. For a single fluid we can use $u_a^{(+)}$ and $u_a^{(-)}$ defined in eq.~\eqref{eq:ic}. We then have
\[
g_{ab}(\eta,\eta_{0})=\sum_{\alpha}u_{a}^{(\alpha)}(\eta)\,c_{b}^{(\alpha)}(\eta_{0})\;, \label{prop_1}
\]
where the coefficients $c_{b}^{(\alpha)}$ are chosen such that
\[
\sum_{\alpha}u_{a}^{(\alpha)}(\eta)\,c_{b}^{(\alpha)}(\eta)=\delta_{ab} \;. \label{prop_2}
\]
For an EdS background the explicit form of $g_{ab}$ reads
\[
g_{ab}(\eta,\eta_{0})=\frac{e^{\eta - \eta_{0}}}{5}
\left(
\begin{array}{cc}
 3  & 2  \\
 3  & 2  \end{array}
\right)
+\frac{e^{-\frac{3}{2}(\eta-\eta_{0})}}{5}
\left(
\begin{array}{cc}
 2  & -2 \\
 -3 & 3  
 \end{array}
\right)
\;.
\]

\begin{figure}
\centerline{\epsfig {figure=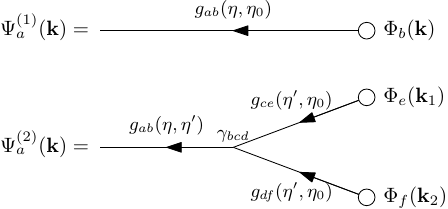,width=8cm}}
\caption{Diagrammatic representation of the series expansion of $\Psi_{a}(\vk)$ up to fourth order in the initial conditions denoted here by $\Phi_a (\vk)$. Time increases along each segment according to the arrow and each segment bears a factor $g_{cd}(\eta_{f}-\eta_{i})$ if $\eta_{i}$ is the initial time and $\eta_{f}$ is the final time. 
At each initial point and each vertex point there is a sum over the component indices; a sum over the incoming wave modes is also implicit and, finally, the time coordinate of the vertex points is integrated from $\eta_0$ to the final time $\eta$ according to the time ordering of each diagram. 
}
\label{DiagramPsiExpansion}
\end{figure}

The linear propagator is useful to formally write the solution for $\Psi_a$ in integral form. Indeed, using eqs.~\eqref{Psi_def} and \eqref{Theta+-},  the equations of motion \eqref{Cont1} and \eqref{Eul1} can be written as \cite{2001NYASA.927...13S}
\[
\begin{split}
&\Psi_{a}(\vk,\eta)=g_{ab}(\eta,\eta_{0})\Psi_{b}(\vk,\eta_0)
\\
&+\int_{\eta_{0}}^{\eta}\dd\eta'g_{ab}(\eta,\eta')\gamma_{bde}(\vk,\vk_{1},\vk_{2})
\Psi_{d}(\vk_{1},\eta')\Psi_{e}(\vk_{2},\eta') \;. \label{sol_g}
\end{split}
\]
As illustrated in Fig.~\ref{DiagramPsiExpansion}, this equation has a diagrammatic representation in the RPT context  \cite{2006PhRvD..73f3519C}.

Another important quantity introduced in the RPT approach is the nonlinear multi-point propagator. More precisely, the $(n+1)$-point propagator $\Gamma^{(n)}_{ab_{1}\dots b_{n}}$ is defined by
\[
\begin{split}
&\left\langle \frac{\partial^n \Psi_{a}(\vk,\eta)}{\partial \Psi_{b_{1}}(\vk_{1},\eta_0)\dots \partial \Psi_{b_{n}}(\vk_{n},\eta_0)}\right\rangle\equiv\\
& \Dirac(\vk - \sum_i^n \vk_i ) \Gamma^{(n)}_{ab_{1}\dots b_{n}}(\vk_{1},\dots,\vk_{n};\eta,\eta_0)\label{GammaAllDef} \;.
\end{split}
\]
Propagators represent the way the $\Psi_a$'s respond to an infinitesimal change of the modes at an earlier time and they are important in the construction of multi-point spectra \cite{2008PhRvD..78j3521B,2010PhRvD..82h3507B}.

In the large-$k$ limit (to be better specified below) these propagators enjoy a remarkable property. Indeed, in \cite{2006PhRvD..73f3520C} it has been shown that in this limit and for Gaussian initial conditions, the nonlinear 2-point propagator $G_{ab} \equiv \Gamma^{(1)}_{ab}$ has a simple expression
\[
G_{ab} (k;\eta,\eta_0) = g_{ab}(\eta,\eta_0) \ \exp\left(-k^2\sigma^2_\displ (e^{\eta}-e^{\eta_0})^2 /2\right)\;,
\label{Gresummed}
\]
where $\sigma^2_\displ$ is the variance of the initial displacement field. Note that the linear propagator $g_{ab}$ is simply the tree level analog of $G_{ab}$. This result has been generalized to $(n+1)$-point propagators with $n \geq 2$ in \cite{2008PhRvD..78j3521B}, where it has been shown that
\[
\Gamma^{(n)}_{ab_{1}\dots b_{n}} = \Gamma^{(n){\rm -tree}}_{ab_{1}\dots b_{n}}  \ \exp\left(-k^2\sigma^2_\displ (e^{\eta}-e^{\eta_0})^2 /2\right)\;,\label{Gammaresummed}
\]
where $\Gamma^{(n)-{\rm tree}}$ is the corresponding propagator computed at tree level.

\begin{figure}
\centerline{\epsfig {figure=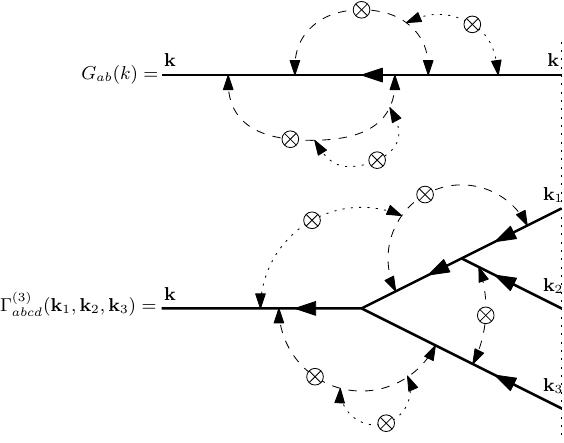,width=8cm}}
\caption{Example of diagrams contributing to $G_{ab}(k)$ (top) and $\Gamma^{(3)}_{abcd}(\vk,\vk_{1},\vk_{2},\vk_{3})$ (bottom).  The dominant contribution after resuming all possible configurations is expected to come from those diagrams where all loops are directly connected to the principal line (top) or principal tree (bottom). The principal line and tree are drawn with a thick solid line. A symbol $\otimes$ denotes a power spectrum evaluated at initial time $\eta_\init$. The dominant loops are those drawn by dashed lines, while the sub-dominant loops are those in dotted lines.}
\label{PrincipalTrees}
\end{figure}

The exponentiation in eqs.~\eqref{Gresummed} and \eqref{Gammaresummed} has been obtained in \cite{2006PhRvD..73f3519C,2006PhRvD..73f3520C} by summing up an infinite number of diagrams thought to dominate in the large-$k$ limit. In order to identify which diagrams dominate in this limit, the concept of \textsl{principal line} and its generalization for the $(n+1)$-point propagators, the \textsl{principal tree}, have been introduced. In \cite{2006PhRvD..73f3520C}  it has been shown that each diagram contributing to the nonlinear propagator $G_{ab}(k;\eta,\eta_0)$ always contains a unique line that goes from some time $\eta_0$ (symbolized by the vertical dotted line) to a final time $\eta$. To this line may be attached loops containing power spectra evaluated at an initial time $\eta_\init$. This is illustrated in Fig.~\ref{PrincipalTrees}, upper panel. The principal line is the unique way to go from $\eta_0$ to $\eta$ without crossing an initial power spectrum $\otimes$, thus moving always in the direction of increasing time. 
Similarly, for each diagram contributing to $\Gamma^{(n)}_{a b_1\dots b_{n}} $ there always exists a unique tree with $n$ branches, the \textsl{principal tree}, that joins $\eta_0$
to $\eta$ (see bottom diagram of Fig.~\ref{PrincipalTrees})  \cite{2008PhRvD..78j3521B}. 

We can now specify under which assumption the relations \eqref{Gresummed} and \eqref{Gammaresummed} have been derived. These are:
\begin{itemize}
\item The multi-point propagators are dominated by those diagrams in which every loop is directly connected to the principal tree.
\item The diagrams are computed and summed up in the limit where the incoming wave modes $q_{i}$ are soft, i.e.~$q_{i}\ll k$.
\end{itemize}
As we will show below, the eikonal approximation corresponds exactly to the last assumption. It can incorporate the first one if necessary.

\subsection{Resumming the 2-point propagator with the eikonal approximation}

In \cite{2010PhRvD..82h3507B} it has been shown that eqs.~\eqref{Gresummed} and \eqref{Gammaresummed} can be obtained irrespective of the diagrammatic representations and of the nature of the initial conditions. Indeed, the nonlinear fluid equations contain nonlinear terms that couple short and long-wavelength modes. The eikonal approximation corresponds to study the effect of very long-wavelength modes $q$ on the dynamics of a given short-wavelength mode $k$, in the limit of $q \ll k$. In this limit, space variations of the long-wavelength modes are tiny with respect to the mode $k$, and the long modes can be treated as an external random background. If we neglect the mode couplings between short scales, the nonlinear fluid equations can be rewritten as linear equations embedded in an external random medium.

Let us be more explicit here. Coupling terms are given by a convolution of fields taken at wave modes $\vk_{1}$ and $\vk_{2}$ such that $\vk=\vk_{1}+\vk_{2}$. These nonlinear terms can be split into two different contributions: the one coming from coupling two modes of very different amplitudes, $k_{1}\ll k_{2}$ or $k_{2}\ll k_{1}$, and the one coming from coupling two modes of  comparable amplitudes. In the first case, the small wave modes ought to be much smaller than $\vk$ itself. Let us denote these small modes by $\vq$. In the limit of $q \ll k$, the equations of motion \eqref{EOM} can be rewritten as
\[
\begin{split}
&\frac{\partial}{\partial \eta}\Psi_{a}(\vk)+\Omega_{ab}\Psi_{b}(\vk)=\Xi_{ab}(\vk)\Psi_{b}(\vk) \\
&+
\left[ \gamma_{abc}(\vk,\vk_{1},\vk_{2})\Psi_{b}(\vk_{1})\Psi_{c}(\vk_{2}) \right]_{\rm \cal H}\label{eikonalNL} \;,
\end{split}
\]
with
\[
\Xi_{ab}(\vk,\eta)\equiv  2 \int_{{\cal S}}\dd^{3}\vq \; 
\egamma_{abc}(\vk,\vk,\vq)
\Psi_{c}(\vq,\eta)  \;. \label{Xi_def}
\]
The key point is that in eq. (\ref{Xi_def}) the domain of integration is restricted to the soft momenta, for which $q\ll k$. Conversely, on the right-hand side of eq.~\eqref{eikonalNL} the convolution is done excluding the soft domain, i.e.~it is over hard modes or modes of comparable size. 

In the limit of separation of scales, $\Xi_{ab}$ is a random quantity which depends on the initial conditions. Using eqs.~\eqref{alpha} and \eqref{beta}, for $q\ll k$ the leading expression of the coupling matrix is obtained with the following limit values
$\alpha(\vq,\vk)\approx (\vq \cdot \vk)/q^{2}$, $\alpha(\vk,\vq)\approx0$, 
$\beta(\vq,\vk) =\beta(\vk,\vq)\approx (\vq \cdot \vk)/(2q^{2})$.
Thus, $\egamma_{abc}$ in eq.~\eqref{Xi_def} simplifies and $\Xi_{ab}$ becomes proportional to the identity, with
\[
\begin{split}
\Xi_{ab} (\vk,\eta)  &= \Xi(\vk,\eta) \; \delta_{ab} \;, \\
\Xi(\vk,\eta) &\equiv \int_{{\cal S}}\dd^{3}\vq \; \frac{\vk \cdot\vq}{q^{2}} \; 
 \Theta (\vq,\eta)  \;. \label{Xi_2}
\end{split}
\]
Note that only the velocity field $\Theta$ (and not the density field $\delta$) contributes to $\Xi_{ab}$. Furthermore, as $\Theta(\vx,\eta)$ is real $\Theta(-\vq)=\Theta^*(\vq)$ and thus $\Xi$ is  purely imaginary.

In eq.~\eqref{eikonalNL} we have reabsorbed the effect of the nonlinear coupling with long-wavelength modes in the linear term $\Xi_{ab} \Psi_b$. The solution to this equation can be given in terms of the resummed propagator $\xi_{ab}(\vk,\eta,\eta')$ \footnote{In this paper we will indistinguishably use the term of propagator for both $\xi_{ab}$ and $G_{ab}$ although the latter is the ensemble average of the former.} satisfying the equation
\[
\left(\frac{\partial}{\partial \eta}-\Xi(\vk,\eta)\right)\xi_{ab}(\vk,\eta,\eta')+\Omega_{ac}(\eta)\xi_{cb}(\vk,\eta,\eta')=0\;, \label{xiEOM}
\]
and reads 
\[
\begin{split}
&\Psi_{a}(\vk,\eta)=\xi_{ab}(\eta,\eta_{0})\Psi_{b}(\vk,\eta_0)\\
&+\int_{\eta_{0}}^{\eta}\dd\eta' \xi_{ab}(\eta,\eta')  \left[ \gamma_{bde}  (\vk,\vk_{1},\vk_{2})
\Psi_{d}(\vk_{1},\eta')\Psi_{e}(\vk_{2},\eta') \right]_{\cal H} \;, \label{prop_xi}
\end{split}  
\]
where in the last line  the convolution is done on the hard domain ${\cal H}$.

In the case of a single fluid, as discussed here, eq.~\eqref{xiEOM} can be easily solved. Taking into account the boundary condition $\xi_{ab}(\vk,\eta,\eta)=\delta_{ab}$, one obtains
\[
\xi_{ab}(\vk,\eta,\eta_0)=g_{ab}(\eta,\eta_0)\exp\left(\int^{\eta}_{\eta_0}\dd\eta'\ \Xi(\vk,\eta')\right)\;. \label{resummed_1}
\]
The argument of the exponential is the time integral of the velocity projected along the direction $\vk$, i.e.~the \textsl{displacement} component along $\vk$. Note that in their original calculation, Crocce and Scoccimarro assumed that the incoming modes in the soft (i.e.~large-scale) lines were in the linear and growing regime. Here we need not make this assumption. Equation~\eqref{resummed_1} is valid irrespective of the fact that the incoming modes in $\Xi$ are in the growing mode or not. 

There is another important aspect of eq.~\eqref{resummed_1}. Since $\Xi(\vk,\eta)$ is a purely imaginary number, the soft modes change only the phase of the small-scale modes but not their amplitude. Such an effect will then have no impact on the equal-time power spectra. However, it has some impact on the amplitude of the propagators. Indeed, the phase change inevitably damps the correlation between modes at different times. This effect is at the heart of the regularization scheme used by approaches such as RPT.

To illustrate this last point, let us  see how one can recover eq.~\eqref{Gresummed} using the solution \eqref{prop_xi} and the resummed propagator \eqref{resummed_1} derived with the eikonal approximation. Deriving eq.~\eqref{prop_xi} with respect to an initial field $\Psi_b(\vk,\eta_0)$ as in eq.~\eqref{GammaAllDef}, and taking the ensemble average one finds
\[
G_{ab}(k,\eta,\eta_0)=\langle\xi_{ab}(\vk,\eta,\eta_0)\rangle_{\Xi}\;. \label{bff}
\]
The nonlinear 2-point propagator $G_{ab}$ is given by the ensemble average of $\xi_{ab}(\vk,\eta,\eta_0)$ over the realizations of $\Xi(\vk)$. In general, the expression of the nonlinear propagator introduces the cumulant generating functions
of $\Xi$. Indeed, using \eqref{resummed_1} eq.~\eqref{bff} yields
\[
G_{ab}(k,\eta,\eta_0)=g_{ab}(\eta,\eta_0)\exp\left(\sum_{p=2}^{\infty}\frac{c_{p}}{p!}\right)\;, \label{G_cumulants}
\]
where $c_{p}$ is the $p$-order cumulant of the field $  \int^{\eta}_{\eta_0}\dd\eta'\,\Xi(\vk,\eta')$ and for symmetry reasons the sum is restricted to even values of $p$, thus ensuring that the nonlinear propagators are real.

For Gaussian initial conditions and assuming that at late time the long-wavelength $\Xi$ is in the linear growing mode, cumulants with $p>2$ in eq.~\eqref{G_cumulants} vanish and we are left with only $c_2$. This is given by 
\[
c_2(k) = \int_{\eta_0}^{\eta} \dd \eta' \dd \eta'' \langle \Xi(\vk, \eta') \Xi(\vk,\eta'') \rangle\;. \label{c2}
\]
Then, exploiting the time dependence of the linear growing mode, $\Theta \propto D_+ = e^{\eta-\eta_\init}$, and using eq.~\eqref{Xi_2},  we have
\[
\Xi(\vk, \eta) = D_+ (\eta)  \int_{\cal S}  \dd^3 \vq \frac{\vk \cdot \vq}{q^2} \Theta(\vq,\eta_\init)\;. \label{growing_Theta}
\]
Plugging this expression in eq.~\eqref{c2}, we can express $c_2$ in terms of the initial power spectrum $P_\init(q)$, defined by 
\[
\langle \Theta(\vq,\eta_\init)  \Theta(\vq',\eta_\init) \rangle \equiv \Dirac (\vq + \vq') P_\init(q)\;.
\]
Indeed, we have
\[
c_{2}(k)=-k^{2}\sigma^{2}_\displ \big(e^{\eta-\eta_{\init}}-e^{\eta_0-\eta_\init}\big)^2 \;,  \label{c2_sf}
\]
where $\sigma^{2}_\displ$
gives the variance of the displacement field defined as \cite{2006PhRvD..73f3519C,2006PhRvD..73f3520C}
\[\label{sigmadsq}
\sigma^2_\displ \equiv    \frac13 \int_{\cal S} \dd^3 \vq \frac{P_\init(q)}{q^2}   \;. 
\]
At this stage $\sigma_\displ^2$ depends on the domain of integration and hence on $k$. The standard RPT results are obtained by taking the value of $\sigma_\displ^2$ in the large-$k$ limit. We will comment on this assumption in the conclusion. Then, setting here and in the following $\eta_{\init}=0$ for convenience,  from eq.~\eqref{G_cumulants} we  recover eq.~\eqref{Gresummed}, 
\[
G_{ab}(k,\eta,\eta_0)=g_{ab}(\eta,\eta_0)\exp\left(-k^{2}\sigma^{2}_\displ (e^{\eta}-e^{\eta_0})^2/2\right) \;.  \label{ea1}
\]

\subsection{Higher-order propagators}

Although the focus of this paper is on the 2-point propagator, let us comment on the use of the eikonal approximation, in particular of eq.~\eqref{prop_xi}, in investigating the resummation of higher-order propagators. 

The computation of the nonlinear $3$-point propagator proceeds by replacing $\Psi_d$ and $\Psi_e$ in the second line of eq.~\eqref{prop_xi} by the linear solution given by the first line of this equation. Deriving twice with respect to the initial field  yields 
\[
\begin{split}
&\frac{\partial^2 \Psi_a (\vk, \eta)}{\partial \Psi_b (\vk_1, \eta_0) \partial \Psi_c (\vk_2, \eta_0)}=  \int_{\eta_0}^{\eta} \dd \eta' \xi_{ad} (\vk;\eta,\eta')   \\ 
&\times \left[ \gamma_{def}  (\vk,\vk_{1},\vk_{2}) \xi_{eb}(\vk_1;\eta', \eta_0)\xi_{fc}(\vk_2;\eta', \eta_0) \right]_{\cal H}\;. \label{(43)}
\end{split}
\]
This is the same formal expression as for the naked theory except that here the convolution is restricted to the hard-mode domain ${\cal H}$. Note that the coupling vertex between modes with hard momenta in the second line is not affected by the use of the eikonal approximation: It is identical to the one of the naked theory. Moreover, it is remarkable to see that, using the form given by eq.~\eqref{resummed_1}, the exponential terms factor out of the time integral and their arguments sum up to give
\[
\begin{split}
& \frac{\partial^2 \Psi_a (\vk, \eta)}{\partial \Psi_b (\vk_1, \eta_0) \partial \Psi_c (\vk_2, \eta_0)}=   \exp \left(\int^{\eta}_{\eta_0}\dd\eta'\ \Xi(\vk,\eta')\right)   \\ 
&\times \! \int_{\eta_0}^{\eta} \! \dd \eta' g_{ad} (\eta,\eta') \left[ \gamma_{def}  (\vk,\vk_{1},\vk_{2})  g_{eb}(\eta', \eta_0)g_{fc}(\eta', \eta_0) \right]_{\cal H}. \label{(44)}
\end{split}
\]
Finally, taking the ensemble average and using the definition of multi-point propagators, eq.~\eqref{GammaAllDef}, one obtains in the Gaussian case
\[
\Gamma_{abc}^{(2)}(\eta,\eta_0) =
\Gamma^{(2)-{\rm tree}}_{abc}(\eta,\eta_0) \ \exp 
\left(- k^2 \sigma_\displ^2 (e^{\eta}-e^{\eta_0})^2 \right/2) \;.
\]

This result can be generalized to propagators of any higher order. The formal expressions of the resummed trees computed in the eikonal approximation are obtained from those computed in the naked theory by simply changing the propagators from $g_{ab}$ to $\xi_{ab}$. Then, for each pair of merging branches with equal initial time, one can factor out the phase similarly to what is done when going from eq.~\eqref{(43)} to \eqref{(44)}. Finally, this leaves an overall factor
$\exp\left(\int^{\eta}_{\eta_0}\dd\eta'\ \Xi(\vk,\eta')\right)$,
which can be factorized out, recovering eq.~\eqref{Gammaresummed}. The eikonal approximation explicitly shows  how the results  \cite{2008PhRvD..78j3521B,2010PhRvD..82h3507B} can be recovered and generalized to any time-dependent large-scale wave mode.

\section{Multi-fluids}
\label{Multi-fluids}

In this section we explore the case where the universe is filled with several non-interacting pressureless fluids and show how the eikonal  approximation can be implemented in this case. Due to the gravitational coupling and the expansion, at late time such a system becomes indistinguishable from a single-fluid component. However, during its evolution it can behave very differently from a single perfect fluid depending on the initial conditions. 

\subsection{The equations of motion}

Denoting each fluid by a subscript $\bb$,
the continuity equation reads, for each fluid,
\[
\frac{\partial}{\partial t} \delta_{\bb}+\frac1a \left((1+\delta_{\bb})u_{\bb}^{i}\right)_{,i}=0\;,
\]
while the Euler equation reads
\[
\frac{\partial}{\partial t}u_{\bb}^{i}+H u_{\bb}^{i}+\frac{1}{a}u_{\bb}^{j}u_{\bb,j}^{i}=-\frac{1}{a}\phi_{,i}\;.
\]
The Poisson equation \eqref{poisson}, where now $\delta$ is the density contrast of the total fluid energy density, i.e. 
\[
\rho_{\rm m} \equiv \sum_\bb \rho_{\bb} \equiv  (1+\delta_{\rm m} ) \rhob_{\rm m} \;,
\]
allows to close the system. This introduces couplings between the fluids.

In Fourier space, the equations of motion become now
\begin{widetext}
\begin{align}
\frac{1}{H}\frac{\partial}{\partial t}\delta_{\bb}(\vk)+\theta_{\bb}(\vk)&=- \alpha(\vk_{1},\vk_{2})\theta_{\bb}(\vk_{1})\delta_{\bb}(\vk_{2}) \;,
\label{Cont2}\\
\frac{1}{H}\frac{\partial}{\partial t}\theta_{\bb}(\vk)+\frac{1}{H} \frac{\dd \ln (a^2 H)}{\dd t} \theta_{\bb}(\vk)+\frac{3}{2}\Omega_{\rm m}\delta_{\rm m} (\vk)&= - \beta(\vk_{1},\vk_{2})\theta_{\bb}(\vk_{1})\theta_{\bb}(\vk_{2})\label{Eul2} \;,
\end{align}
\end{widetext}
where $\theta_{\bb}$ is the dimensionless divergence of the velocity field of the fluid $\bb$ and
$\Omega_{\rm m}$ is the reduced total density of the pressureless fluids. The coupling between the fluids is only due to the term $\delta_{\rm m} $ appearing in the Euler equation.

Before studying these equations let us discuss the equations for the total fluid. As we are describing here a collection of pressureless particles, it is tempting to write down the equations of motion for the total fluid. The continuity equation is simply identical to eq.~\eqref{continuity}. The total fluid velocity $u^i$ is defined by
\[
u^{i} \equiv \sum_\bb f_\alpha u_{\bb}^{i} \;,
\]
where
$f_\bb \equiv  {\rho_{\bb}}/{\rho_{\rm m}}$, and its evolution equation reads
\[
\frac{\partial}{\partial t}u^{i}+H u^{i}+\frac{1}{a}u^{j}u^i_{,j} =
-\frac{1}{a}\phi_{,i}-\frac{1}{a\rho_{\rm m}}\left(\rho_{\rm m}\sigma^{ij}\right)_{,j}\;,
\]
where 
and $\sigma_{ij}$ is the velocity dispersion of the mean fluid, given by
\[
 \sigma^{ij} \equiv  \sum_\bb f_{\bb}  u_{\bb}^{i}u_{\bb}^{j}-u^{i}u^{j} \;. \label{sigmaij}
\]
Thus, due to the multi-fluid nature of the system, the Euler equation contains an anisotropic stress term. One can write down an equation of motion for this term, but this will involve higher moments of the fluid distribution and so on. Thus, the complete description of the total fluid at nonlinear order requires an infinite hierarchy of equations in the moments of the fluid.
Another consequence of this expression is that, even though the velocity field of each fluid remains potential, the total velocity field is no longer potential. Indeed, we expect that it develops a rotational part due to the presence of the dissipative term $\sigma^{ij}$ \footnote{We are here in a situation comparable to that encountered in Lagrangian space where the displacement is found to be non-potential at order three and beyond in PT.}.

\subsection{Adiabatic and isodensity modes}

As we did for the single-fluid case, let us study the linear evolution of the multi-fluid system by dropping the right-hand side of eqs.~\eqref{Cont2} and \eqref{Eul2}. Since at linear order there is no anisotropic stress $\sigma^{ij}$, which is second order in the velocities, the two linear solutions \eqref{delta+-} and \eqref{Theta+-} found in the single-fluid case are expected to be also solutions of the linear multi-fluid system. This corresponds to the case where all the fluids start comoving and, as they all follow geodesic motion, remain comoving during their entire evolution. Analogously to the jargon adopted in the physics of the early universe, these solutions correspond to the so-called growing and decaying  \textsl{adiabatic} modes. 
Note that if only these two modes are initially excited, the right-hand side of eq.~\eqref{sigmaij} vanishes and the total fluid is indistinguishable from a pure dark matter fluid.

However, the presence of multiple components gives birth also to \textsl{isocurvature} or rather, given our scales of interest, \textsl{isodensity} modes. To examine their properties, let us turn to the equations describing the multi-component system, eqs.~\eqref{Cont2} and \eqref{Eul2}. In this case it is convenient to introduce a multiplet $\Psi_a$ which generalizes the duplet defined in eq.~\eqref{Psi_def}, i.e.~\cite{2001NYASA.927...13S}
\[
\Psi_{a}=\left(
\delta_{1},
\rtheta_{1},
\delta_{2},
\rtheta_{2},
\dots
\right)^T \;, \label{Psi_def_2}
\]
where $
\rtheta_{\bb} \equiv - {\theta_{\bb}}/{f_{+}(t)}$.
Thus, for $N$ components $\Psi_a$ has $2N$ elements. Equations~\eqref{Cont2} and \eqref{Eul2} can then be rewritten as eq.~\eqref{EOM} where in this case the matrix elements of $\Omega_{ab}$ are given by
\[
\begin{split}
\Omega_{(2p-1)\,(2p)}&=-1\;, \\
\Omega_{(2p)\,(2p)}&=\frac{3}{2}\frac{\Omega_{m}}{f_{+}^{2}}-1\;, \\
\Omega_{(2p)\,(2q-1)}&=-\frac{3}{2}\frac{\Omega_{m}}{f_{+}^{2}}f_{q} \;,
\end{split}
\]
for any integers $p$ and $q$ running from 1 to $N$ and where all the other elements of $\Omega_{ab}$ vanish.
The non-zero elements of the coupling matrix $\gamma_{abc}$ are
\[
\begin{split}
\gamma_{(2p-1)\,(2p-1)\,(2p)}(\vk,\vk_{1},\vk_{2})
&=\frac{\alpha(\vk_{2},\vk_{1})}{2}\;,
\\
\gamma_{(2p-1)\,(2p)\,(2p-1)}(\vk,\vk_{1},\vk_{2})
&=\frac{\alpha(\vk_{1},\vk_{2})}{2}\;,
\\
\gamma_{(2p)\,(2p)\,(2p)}(\vk,\vk_{1},\vk_{2})
&=\beta(\vk_{1},\vk_{2}) \;,
\end{split}
\]
for any integer $p$. Note that there are no explicit couplings between different species in the $\gamma_{abc}$-matrices.

The isodensity modes are obtained under the constraint that the total density contrast vanishes, i.e.~$\delta=0$. Since the evolution equations decouple under this constraint, the time dependence of these modes can be easily inferred. One solution is given by
\[
\begin{split}
\rtheta_{\bb}^{(\ii)}(\eta)
&\propto \exp\left[-\int^{\eta}\dd\eta'\;
\left(\frac{3}{2}\frac{\Omega_{m}}{f_{+}^{2}}-1\right)\right] \;,
\\
\delta_{\bb}^{(\ii)}(\eta)
&=\int^\eta \dd \eta'\; \rtheta_{\bb}^{(\ii)}(\eta') \;,
\end{split}
\]
with
\[
\sum_\bb f_\bb \rtheta_{\bb}^{(\ii)}=0 \;, \label{Theta_sum_i}
\]
which automatically ensures that $\sum_{\bb}f_{\bb}\delta_{\bb}^{(\ii)}=0$. Note that as  $\Omega_{m}/f_{+}^2$ departs little from the value taken in an EdS cosmology, i.e.~$\Omega_{m}/f_{+}^2=1$, the isodensity modes are expected to depart very weakly from
\[
\rtheta_{\bb}^{(\ii)}(\eta)\propto \exp(-\eta/2)\; ,\qquad \delta_{\bb}^{(\ii)}(\eta)=-2\rtheta_{\bb}^{(\ii)}(\eta) \;.\label{i_mode}
\]
A second set of isodensity modes is given by
\[
\rtheta_{\bb}^{(\ci)}(\eta)=0\; ,\qquad \delta_{\bb}^{(\ci)}(\eta)=\hbox{Constant}\;, \label{ci_mode}
\]
again with
\[
\sum_{\bb}\ f_{\bb}\,\delta_{\bb}^{(\ci)}=0 \;.
\]

To be specific, let us concentrate now on the case of two fluids and assume an EdS background. In this case the growing and decaying solutions are proportional, respectively, to 
\[
\begin{split}
u_{a}^{(+)} & \propto\left(1,1,1,1\right)^T \;,\\
u_{a}^{(-)} & \propto\left(1,-3/2,1,-3/2\right)^T \;.
\end{split}
\]
Moreover, the isodensity modes are proportional to
\[
\begin{split}
u_{a}^{(\ii)} & \propto\left(-2 f_2 ,f_2,2f_1,-f_1\right)^T \;, \\
u_{a}^{(\ci)} & \propto\left(f_2 ,0,-f_1,0\right)^T \;.
\end{split}
\]

We are then in position to write down the linear propagator $g_{ab}(\eta,\eta_0)$ satisfying eqs.~\eqref{prop_1} with \eqref{prop_2}. For two fluids and an EdS background it reads \cite{2010PhRvD..81b3524S} 
\begin{widetext}
\[
\begin{split}
g_{ab}(\eta,\eta_0)= \ &\frac{e^{\eta-\eta_0}}{5}
\left(
\begin{array}{cccc}
 3 f_1 & 2 f_1 & 3 f_2 & 2 f_2 \\
 3 f_1 & 2 f_1 & 3 f_2 & 2 f_2 \\
 3 f_1 & 2 f_1 & 3 f_2 & 2 f_2 \\
 3 f_1 & 2 f_1 & 3 f_2 & 2 f_2
\end{array}
\right)
+\frac{e^{-\frac32(\eta-\eta_0)}}{5}
\left(
\begin{array}{cccc}
 2 f_1 & -2 f_1 & 2 f_2 & -2 f_2 \\
 -3 f_1 & 3 f_1 & -3 f_2 & 3 f_2 \\
 2 f_1 & -2 f_1 & 2 f_2 & -2 f_2 \\
 -3 f_1 & 3 f_1 & -3 f_2 & 3 f_2 
\end{array}
\right) \\
&+
e^{-\frac12(\eta-\eta_0)}
\left(
\begin{array}{cccc}
 0 & -2 f_2 & 0 & 2 f_2 \\
 0 & f_2 & 0 & -f_2 \\
 0 & 2 f_1 & 0 & -2 f_1 \\
 0 & -f_1 & 0 & f_1
\end{array}
\right)
+
\left(
\begin{array}{cccc}
 f_2 & 2 f_2 & -f_2 & -2 f_2 \\
 0 & 0 & 0 & 0 \\
 -f_1 & -2 f_1 & f_1 & 2 f_1 \\
 0 & 0 & 0 & 0
\end{array}
\right) \;. \label{linear_g}
\end{split}
\]
\end{widetext}
In the following we explore how this propagator is changed by the coupling with the long-wavelength modes in the eikonal approximation.

\subsection{Resummation of the propagator with the eikonal approximation}

Let us study the resummed propagator in the presence of more than one fluid. For simplicity, we will restrict the study to the two-fluid case and an EdS background.

The eikonal equation, eq.~\eqref{eikonalNL} with \eqref{Xi_def}, also holds in the multi-fluid case. However, in this case $\Xi_{ab}$ is given by a sum of adiabatic contributions, for which the fluid displacements are the same, and isodensity contributions, for which their weighted sum vanishes, i.e.,
\[
\Xi_{ab}(\vk,\eta)=\Xi^{({\rm ad})} (\vk,\eta)  \delta_{ab}+\Xi^{(\ii)}_{ab}(\vk,\eta)\;,
\label{Xi_3}
\]
where $\Xi_{ab}^{(\ii)}$ takes the form
\[
\Xi_{ab}^{(\ii)} = \Xi^{(\ii)} \; h_{ab} \;, \quad h_{ab} \equiv
\left(
\begin{array}{cccc}
 f_2 & 0 & 0 & 0 \\
 0 & f_2 & 0 &  0 \\
 0 & 0 & -f_1 & 0 \\
 0 & 0 & 0 & - f_1
\end{array}
\right) \;. \label{iso_prop}
\]

If we assume $\Xi_{ab}$ to be in the linear regime, then
\[
\Xi^{({\rm ad})}(\vk,\eta) \equiv \int_{\cal S} \dd^3 \vq \frac{\vk \cdot \vq }{q^2} \big( \Theta^{(+)} (\vq,\eta) + \Theta^{(-)} (\vq,\eta)\big) \;,
\]
where $\Theta^{(+)}$ and $\Theta^{(-)}$ are, respectively, the growing and decaying adiabatic modes of the long-wavelength displacement field. The isodensity contribution $\Xi^{(\ii)}_{ab}$ contains the decaying isodensity mode given in eq.~\eqref{i_mode}, so that it reads
\[
\Xi^{(\ii)} \equiv \frac{1}{f_2} \int_{\cal S} \dd^3 \vq \frac{\vk \cdot \vq }{q^2} \Theta_1^{(\ii)}  = - \frac{1}{f_1} \int_{\cal S} \dd^3 \vq \frac{\vk \cdot \vq }{q^2} \Theta_2^{(\ii)}  \;.
\]
Note that, because of eq.~\eqref{ci_mode}, the constant isodensity mode does not contribute
to $\Xi_{ab}$.

We are now interested in computing the resummed propagator in the eikonal approximation under the modulation of the long-wavelength modes in eq.~\eqref{Xi_3}. As the adiabatic modes in $\Xi_{ab}$ are proportional to the identity, their effect can be incorporated in exactly the same manner as in the single-fluid case. The adiabatic modes will contribute to the resummed propagator by a multiplicative factor of the exponential of the adiabatic displacement field, as in eq.~\eqref{resummed_1},
\[
\begin{split}
\xi_{ab}(\vk;\eta,\eta_0) = & \; \xi_{ab}(\vk; \eta,\eta_0; \Xi^{(\rm ad)}=0) \\&\times \exp \left( \int^{\eta}_{\eta_0}\dd\eta' \Xi^{({\rm ad})}(\vk,\eta') \right)\;.
\end{split}
\]
Note again that, as in the single-fluid case, the soft adiabatic modes induce a phase change but do not affect the amplitude of the small-scale modes.

\begin{figure}
\centerline{\epsfig {figure=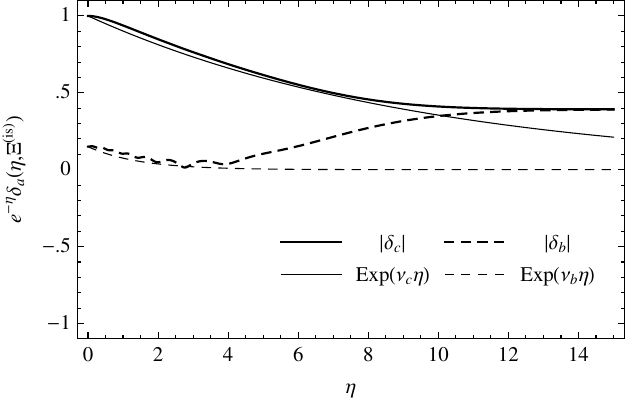,width=8.5cm}}
\vspace{4mm}
\centerline{\epsfig {figure=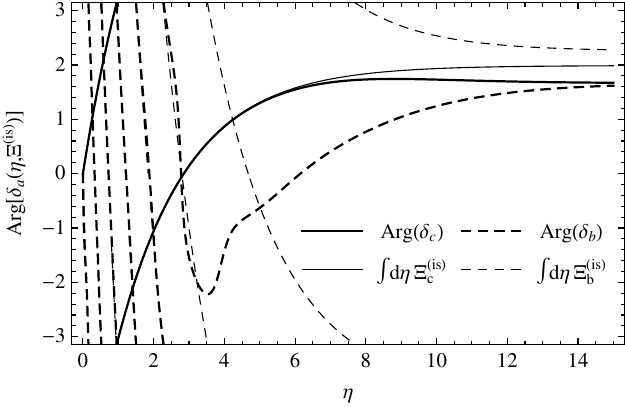,width=8.5cm}}
\caption{Evolution of the amplitudes (upper panel) and phases (lower panel) of the resummed density modes with $\Xi^{(\rm ad)}=0$, i.e.~$\delta_\cdm (\eta,  \Xi^{(\ii)})$ and $\delta_\ba(\eta,  \Xi^{(\ii)})$ as defined in eq.~\eqref{iso_dd} for CDM (thick solid line) and baryons (thick dashed line). The initial conditions are chosen such that $\Psi_{a}(\eta_\init)=(1,1,0.15,0.15)^{T}$ and $|\Xi^{(\rm is)}(\eta_\init)|=25$. Thin lines represent the analytic solutions at early times of eq.~\eqref{sol_et}.}
\label{IsoEvol}
\end{figure}
Including the isodensity mode in the resummed propagator proved difficult. We have not been able to find a closed analytic form for it. Thus, we have to rely either on numerical studies or on perturbative calculations. As an example, in Fig.~\ref{IsoEvol} we show the effect of the soft isodensity mode on the small-scale modes, by plotting the evolution of the resummed CDM and baryon density modes with $\Xi^{(\rm ad)}=0$, i.e.,
\[
\begin{split}
\delta_\cdm(\eta,  \Xi^{(\ii)}) &\equiv \xi_{1a}(k; \eta,\eta_\init; \Xi^{(\rm ad)}=0) \Psi_a(\eta_{\init})\;, \\
\delta_\ba(\eta,  \Xi^{(\ii)}) &\equiv \xi_{3a}(k; \eta,\eta_\init; \Xi^{(\rm ad)}=0) \Psi_a(\eta_{\init})\;, \label{iso_dd}
\end{split}
\]
normalized to the growing mode $e^{\eta}$. Initial conditions are chosen such that $\Psi_a (\eta_\init)= (1,1,0.15,0.15)^T$ and we have taken $|\Xi^{(\rm is)}(k,\eta_{\init})|=25$. At early time the CDM and baryon density mode grow slower than $e^\eta$ (upper panel) and since $|\Xi^{(\rm is)}| \gg 1$ the phases of the two modes rapidly evolve (lower panel). At late time the phases are fixed and the density modes evolve according to the standard adiabatic growing mode. Let us study these two limiting behaviors.

\subsubsection{Early-time behavior}

As $\Xi^{(\ii)}$ is a decaying mode, it can become arbitrarily large at early time. Let us consider a mode $k$ for which initially $|\Xi^{(\ii)}| \gg 1$. This means that $\vk \cdot \vv_{\cal S}$, i.e.~the displacement field of the soft modes along $\vk$, is much larger than the Hubble flow. In other words, the time scale of the motion of the large-scale modes is much shorter than the time-scale of growth of the small-scale ones, set by the Hubble time.

We can grasp the nature of the early-time evolution by making the following change of variable,
\[
\tPsi_{a}(\eta)=\Psi_{a} (\eta) \exp\left(- \int^{\eta}_{\eta_0} \dd \eta' \; \Xi_{aa} (\eta')\right)\;,\label{tildepsi_vc}
\]
where there is no summation over $a$ in $\Xi_{aa}$. In this case the first line (i.e.~the large-$k$ part) of eq.~\eqref{eikonalNL} can be rewritten in terms of $\tilde \Psi_a$ as
\[
\frac{\partial}{\partial\eta}\tPsi_{a}(\eta)+\tilde \Omega_{ab}\tPsi_{b}(\eta)=\tilde \Xi_{ab}(\eta) \tilde \Psi_b(\eta)\;, \label{tildepsi}
\]
where 
\[
\tilde \Omega_{ab} \equiv \left(\begin{array}{cccc}
 0 & -1 & 0 & 0 \\
 -{3} f_1/{2} & {1}/{2}  & 0 & 0 \\
0 & 0 & 0 & -1 \\
0& 0 & -{3}f_2 /{2} & {1}/{2}    
 \end{array} \right)\;,
\]
and 
\[
\tilde \Xi_{ab} \equiv
\left(
\begin{array}{cccc}
0 & 0 & 0 & 0 \\
0 & 0 & 3f_2 e^{-i\varphi}/2 & 0 \\
0 & 0 & 0 & 0 \\
3f_1 e^{i\varphi}/2 & 0 & 0 & 0
\end{array}
\right)\;,
\]
with
\[
\varphi (\eta)\equiv - i \int^\eta_{\eta_0} \dd\eta' \Xi^{(\ii)}(\eta') \;.  \label{phase}
\]
The two fluids are only coupled through $\tilde \Xi_{ab}$. However, $\Xi^{(\ii)}$ in eq.~\eqref{phase} is purely imaginary: At early time the coupling term contributes to a rapidly changing phase $\varphi$. When the time scale  of these oscillations is much shorter than that of  structure growth, this force term can effectively be neglected and the different species  decouple. Indeed, the velocity difference  (in the direction along $\vk$) between the coherent flows of the two species  is large enough that the short modes of one fluid do not gravitationally see those of the other fluid.

The system we are left with is given by eq.~\eqref{tildepsi} with vanishing right-hand side. For an EdS background the solution of this equation is given by \cite{1980PhRvL..45.1980B}  
\[
\tilde \delta_\bb \propto \exp(\nu^{({\pm})}_\alpha \eta)\;, \label{sol_et}
\]
with
\[
\nu^{({\pm})}_\alpha=\frac{1}{4}\left(-1\pm\sqrt{1+24f_\bb}\right) \;.
\]
The growing solutions in eq.~\eqref{sol_et} explain the  early-time evolution shown in Fig.~\ref{IsoEvol}. At early time, both CDM and baryons grow slower than $e^\eta$ and their phases are dominated by their respective large-scale isocurvature displacement fields, as accounted for by the change of variable \eqref{tildepsi_vc}.

\subsubsection{Late-time behavior}

As the isodensity mode decays, one expects to recover at late time the single-fluid propagator 
$g_{ab}(\eta,\eta_0)\exp \left( \int^{\eta}_{\eta_0}\dd\eta' \Xi^{({\rm ad})}(\eta') \right) $. 
More precisely, we can compute how the propagator $\xi_{ab}$ deviates from the adiabatic one with a perturbative analysis. Indeed, since $\Xi^{(\ii)}_{ab}$ becomes small at late time, one can compute $\xi_{ab}(\vk;\eta,\eta_0)$ perturbatively in $\Xi^{(\ii)}$. 

Solving the first line of eq.~\eqref{eikonalNL} at first order in $\Xi^{(\ii)}$ yields,
\begin{widetext}
\[
\xi_{ad} (\vk, \eta,\eta_0) \approx \exp \left( \int^{\eta}_{\eta_0}\dd\eta' \Xi^{({\rm ad})}(\vk,\eta') \right)  \bigg[ g_{ad}(\eta,\eta_0)  + \int_{\eta_0}^\eta \dd \eta' g_{ab}(\eta,\eta') \Xi_{bc}^{(\ii)} (\vk,\eta') g_{cd} ( \eta',\eta_0) \bigg]\;. 
\]
By plugging in this equation the expressions of the linear propagator $g_{ab}$ from eq.~\eqref{linear_g} and of $\Xi_{ab}^{(\ii)}$ from eq.~\eqref{iso_prop}, and integrating in time yields
\[
\xi_{ab}(\vk;\eta,\eta_0)\approx \left[ g_{ab}(\eta,\eta_0)   +  \Xi^{(\ii)} (\vk, 0) e^{-\eta_0/2}C_{ab}(\eta,\eta_0) \right]\exp \left( \int^{\eta}_{\eta_0}\dd\eta' \Xi^{({\rm ad})}(\vk,\eta') \right) \;, \label{correction1}
\]
with
\[
\begin{split} C_{ab}(\eta,\eta_0) \equiv \ &
 f_1 f_2 \left[
\frac{2 e^{\eta-\eta_0}}{5}
\left(
\begin{array}{cccc}
 1 & 1  & -1 & -1 \\
 1 & 1  & -1 & -1 \\
 1 & 1  & -1 & -1 \\
 1 & 1  & -1 & -1
\end{array}
\right)\right.+
\frac{e^{\frac12(\eta-\eta_0)}}{5}
\left(
\begin{array}{cccc}
 12 & 8 & 12\frac{f_2}{f_1} & 8\frac{f_2}{f_1} \\
 3 & 2 & 3\frac{f_2}{f_1} & 2\frac{f_2}{f_1} \\
 -12 \frac{f_1}{f_2} & -8 \frac{f_1}{f_2} & -12 & -8 \\
 -3 \frac{f_1}{f_2} & -2 \frac{f_1}{f_2} & -3 & -2
\end{array}
\right) \\
&+
2 \left(
\begin{array}{cccc}
 \frac{f_2}{f_1}-3 & 2 \frac{f_2}{f_1} -3 & 1 -3 \frac{f_2}{f_1} & 2 - 3\frac{f_2}{f_1}
   \\
 0 & 0 & 0 & 0 \\
 3\frac{f_1}{f_2}-1 & 3 \frac{f_1}{f_2} -2 & 3 - \frac{f_1}{f_2} & 3 - 2\frac{f_1}{f_2} \\
 0 & 0 & 0 & 0
\end{array}
\right) +
{e^{-\frac12(\eta-\eta_0)}}
\left(
\begin{array}{cccc}
 4 - 2 \frac{f_2}{f_1} & 8-8\frac{f_2}{f_1} & 4\frac{f_2}{f_1} - 2 & 8\frac{f_2}{f_1} -8
    \\
 -2  & 2\frac{f_2}{f_1} -4 & 1-\frac{f_2}{f_1} & 4-2 \frac{f_2}{f_1} \\
 2 - 4 \frac{f_1}{f_2} & 8-8\frac{f_1}{f_2} & 2\frac{f_1}{f_2} - 4 & 8\frac{f_1}{f_2} -8 \\
 \frac{f_1}{f_2}-1 & 2 \frac{f_1}{f_2} -4 & 2 &  4-2\frac{f_1}{f_2}
\end{array}
\right) \\
&+
2 e^{-(\eta-\eta_0)}
\left(
\begin{array}{cccc}
 0 & 2 \frac{f_2}{f_1} - 3 & 0 & 3 - 2\frac{f_2}{f_1} \\
 0 & 3- \frac{f_2}{f_1}  & 0 & \frac{f_2}{f_1}-3 \\
 0 & 2 \frac{f_1}{f_2} - 3 & 0 & 3 - 2\frac{f_1}{f_2} \\
 0 & 3- \frac{f_1}{f_2}  & 0 &  \frac{f_1}{f_2} - 3 
\end{array}
\right) +
\frac{e^{-\frac32(\eta-\eta_0)}}{5}
\left(
\begin{array}{cccc}
 -2 & 8 & 2 & -8 \\
 3 & -12 & -3 & 12 \\
 -2 & 8 & 2 & -8 \\
 3 & -12 & -3 & 12
\end{array}
\right)\\ 
&+
\left.
\frac{2 e^{-2(\eta-\eta_0)}}{5}
\left(
\begin{array}{cccc}
 -1 & 1 & -\frac{f_2}{f_1} & \frac{f_2}{f_1} \\
1 & -1 & \frac{f_2}{f_1} & -\frac{f_2}{f_1} \\
\frac{f_1}{f_2} & -\frac{f_1}{f_2} & 1 & -1 \\
 -\frac{f_1}{f_2} & \frac{f_1}{f_2} & -1 & 1
\end{array}
\right)
\right]\;. \label{correction2}
\end{split}
\]
\end{widetext}
Note that this result is written in terms of $\Xi^{(\ii)}$ taken at the initial time $\eta_\init=0$, so that the time dependence of $\Xi^{(\ii)}$ is included in eq.~\eqref{correction1} and in the square bracket of eq.~\eqref{correction2}. As expected by eq.~\eqref{iso_prop}, the corrections to the propagator due to the isodensity mode are invariant under exchange of $f_1 \leftrightarrow - f_2 $  and $1,2 \leftrightarrow 3,4$ in the matrix indices $a,b$. Furthermore, $C_{ab}(\eta,\eta)=0$.

The final expression of the nonlinear propagator is obtained after the ensemble average of 
$\Xi^{(\ii)}(0) \exp \left( \int^{\eta}_{\eta_0}\dd\eta' \Xi^{({\rm ad})}(\eta') \right) \,$ has been taken. We recall here that the different modes that enter in $\Xi_{ab}$ are not statistically independent. The ensemble average can be written as (see appendix \ref{moments}),
\[
\left\langle \Xi^{(\ii)}(0) e^{ \int^{\eta}_{\eta_0}\dd\eta' \Xi^{({\rm ad})}(\eta')}  \right\rangle 
=
\sum_{p}\frac{x_{1,p-1}}{(p\!-\!1)!}\exp\left(\sum_{q=2}^{\infty}\frac{c_{q}}{q!}\right) \;,\label{cross1}
\]
where 
$c_{q}$ is the $q$-order cumulant of the adiabatic modes and $x_{1,p-1}$ is a $p$-order cross-cumulant defined as
\[
x_{1,p-1} \equiv \left\langle 
\Xi^{(\ii)}(0)\left(\int^{\eta}_{\eta_0}\dd\eta' \Xi^{({\rm ad})}(\eta')\right) 
^{p-1}
\right\rangle_{c} \;.
\]
The explicit values of such coefficients depend on the precise model. For Gaussian initial conditions only $c_2$ and $x_{1,1}$ are non-zero. Then, eq.~\eqref{cross1} can be rewritten as
\[
\left\langle \Xi^{(\ii)}(0) \exp \left( \int^{\eta}_{\eta_0}\dd\eta' \Xi^{({\rm ad})}(\eta') \right)  \right\rangle =  x_{1,1} \exp\left(\frac{c_{2}}{2}\right) \;, \label{ea}
\]
where 
\begin{align}
c_2 &= \int_{\eta_0}^{\eta} \dd \eta' \dd \eta'' \langle \Xi^{({\rm ad})}(\eta') \Xi^{({\rm ad})}(\eta'') \rangle\;, \\
x_{1,1} &= \int_{\eta_0}^{\eta} \dd \eta' \langle \Xi^{(\ii)}(0) \Xi^{({\rm ad})}(\eta') \rangle\;.
\end{align}

At late time $\Xi^{({\rm ad})}$ is dominated by the growing mode. Thus, we can express it as on the right-hand side of eq.~\eqref{growing_Theta} and we can use eq.~\eqref{c2_sf} for $c_2$.
For $x_{1,1}$ we find 
\[
x_{1,1} = -k^2 \sigmax^2 (e^{\eta}-e^{\eta_0})\;, 
\]
where $\sigma^2_{\times}$ is the cross-correlation between the initial isodensity and the adiabatic modes,
\[
\sigmax^2 \equiv  \frac13 \int \dd^3 \vq \frac{C_\init (q)}{q^2} \;,
\]
with $C_\init$ defined by
\[
\left\langle \Theta^{(\ii)}(\vq,0) \Theta^{({\rm ad})}(\vq',0) \right\rangle = \Dirac (\vq + \vq') C_\init(q)\;.
\]
Finally, the ensemble average in eq.~\eqref{ea} can be written as
\[
\left\langle \Xi^{(\ii)} e^{  \int^{\eta}_{\eta_0}\dd\eta' \Xi^{({\rm ad})}(\eta') } \right\rangle=
-k^2\sigmax^2 (e^{\eta}-e^{\eta_0}) e^{-k^2\sigma_\displ^2 (e^{\eta}-e^{\eta_0})^2/2}\;, \label{ea2}
\]
so that the nonlinear propagator reads, at first order,
\begin{widetext}
\[
G_{ab}(\vk;\eta,\eta_0)\approx \left[ g_{ab}(\eta,\eta_0)   -k^2\sigmax^2 (e^{\eta}-e^{\eta_0}) e^{-\eta_0/2} C_{ab}(\eta,\eta_0) \right] e^{-k^2\sigma_\displ^2 (e^{\eta}-e^{\eta_0})^2/2} \;. \label{resummed_prop}
\]
\end{widetext}

It is possible to compute the nonlinear propagator at higher orders in $\Xi^{(\ii)}$. In particular, in appendix~\ref{HOTimeDependence} we derive a recurrence formula for the most growing mode of the resummed propagator, to any order in $\Xi^{(\ii)}$. We are now in the position to illustrate the effect discussed in this section in a practical case, i.e.~the mixture of baryons and  cold dark matter after decoupling.

\section{CDM and baryons after decoupling}
\label{CDMbaryons}

As an application, in this section we consider the case of baryons and CDM particles just after decoupling. This situation is illustrative of the concepts that we introduced in this paper. Here we focus on the behavior of the propagators on scales which are interesting for PT calculations, i.e.~$ k\lesssim 1  h {\rm Mpc}^{-1}$. We will see that for such statistical objects and such scales the impact of isodensity modes is very small. We leave the calculations of power spectra for further studies.

The first step of our analysis is to properly identify the isodensity modes after recombination. We will assume that the \textsl{primordial} (i.e.~before horizon crossing) large-scale perturbations are strictly adiabatic. In this case each fluid component is proportional to the same random field, for instance the primordial curvature perturbation $\zeta(\vk)$. We can assume that at the initial time $\eta_\init =0$ the different fluid variables are in the linear regime. Then, they can be written in terms of the initial linear transfer functions $T_{a}(k,0)$ as
\[
\Psi_{a}(\vk,0)=T_{a}(k,0)\zeta(\vk)\; .
\]
We will use CAMB \cite{Lewis:1999bs} to generate the CDM and baryon initial transfer functions, assuming the following cosmological parameters: $\Omega_\cdm = 0.233$, $\Omega_\ba = 0.0461$, $\Omega_\Lambda = 0.721$, $h=0.700$, $n_s=0.96$, $A_\zeta=2.46 \cdot 10^{-9}$ and massless neutrino species. 

A remark is in order here. Cosmological fluctuations, such as those described by the CAMB code, obey linear general relativistic equations. On large scales, i.e.~on scales comparable with the Hubble radius, these equations may considerably deviate from the Newtonian equations  used in RPT, also at the linear level. Thus, one may worry that the transfer functions generated by CAMB will be affected by these deviations, which  are gauge dependent. However, as shown in appendix~\ref{SH}, for a set of \textsl{pressureless} fluids there exists a choice of variables for which at linear order the relativistic equations  \textsl{exactly} reduce to the Newtonian equations. For the density contrasts of cold dark matter and baryons, this choice corresponds to take the energy density perturbations in a gauge comoving to the total fluid. For the velocity divergences this corresponds to take them in the longitudinal gauge. In the limit where we can neglect radiation energy and momentum, the dynamics of these variables is well described by the Newtonian equations even on super-Hubble scales.

Finally, note that even though $\Omega_\Lambda \neq 0$, we will use the linear propagator derived in sec.~\ref{Multi-fluids} in a EdS universe. Indeed, as explained in \cite{2006PhRvD..73f3520C} most of the cosmological dependence is encoded in the linear growth function $D_+ $ and using the propagators derived for an EdS universe is a very good approximation.

\subsection{The linear modes after decoupling}

\begin{figure}
\begin{center}
      \psfig{file=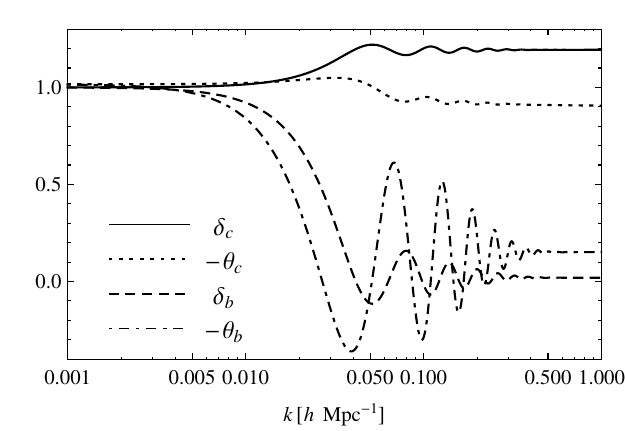,width=8.5cm}
\caption{Shape and amplitude of the transfer functions at $z=900$. The transfer functions are plotted in units of the total density transfer function. From top to bottom we have the CDM density transfer function (continuous line), the CDM velocity transfer function (dotted line), the baryon density transfer function (dashed line) and the baryon velocity transfer function (dotted-dashed line). On super-horizon scales they are all approximately equal, denoting that $f_+ \simeq1$. One can observe that at this high redshift the baryon transfer functions are
highly suppressed.}
\label{SpeciesTransfer}
\end{center}
\end{figure}

In Fig.~\ref{SpeciesTransfer} we show the transfer functions for the different fluid variables normalized to the transfer function of the total matter perturbation $\delta_{\rm m}$. We choose redshift $z=900$ as initial time $\eta_\init =0$. At this redshift the energy density and momentum density of the radiation are still important (of the order of 20\% percents). However, we will neglect their contributions in our treatment. Moreover, since on super-horizon scales the transfer functions are all approximately equal, then $f_+ \simeq 1$ and we will take $\Theta_\alpha = - \theta_\alpha$. Note that, contrary to what has been done in \cite{2010PhRvD..81b3524S}, one cannot consistently assume that the density and velocity transfer functions are the same.

The linear evolution of each mode can be constructed by applying the linear propagator $g_{ab}$ given in eq.~\eqref{linear_g}. In particular, $g_{ab}^{(+)}(\eta,\eta_0)$, $g_{ab}^{(-)}(\eta,\eta_0)$, $g_{ab}^{(\ii)}(\eta,\eta_0)$ and $g_{ab}^{(\ci)}(\eta,\eta_0)$ are the growing and decaying adiabatic, and the decaying and constant isodensity time-dependent projectors, given respectively by the first, second, third and fourth term on the right-hand side of eq.~\eqref{linear_g}. In terms of these projectors one can define the transfer function for each mode as
\[
\begin{split}
T^{(+)} (k,\eta)& \equiv (f_{\cdm},0,f_{\ba},0) \; g_{ab}^{(+)} (\eta,0)\; T_b(k,0)\;, \\
T^{(-)} (k,\eta)& \equiv (f_{\cdm},0,f_{\ba},0)\; g_{ab}^{(-)} (\eta,0)\; T_b(k,0)\;, \\
T^{(\ii)} (k,\eta)& \equiv (0, 1, 0, -1)\;g_{ab}^{(\ii)} (\eta,0)\; T_b(k,0)\;, \\
T^{(\ci)} (k,\eta)& \equiv (1, 0,-1, 0) \;g_{ab}^{(\ci)} (\eta,0)\; T_b(k,0)\;. \label{def_T}
\end{split}
\]
These definitions have been chosen in such a way that 
\[
T_{a} =T^{(+)} u^{(+)}_a +T^{(-)} u^{(-)}_a + T^{(\ii)} u^{(\ii)}_a +T^{(\ci)} u^{(\ci)}_a\;.
\]

\begin{figure}
\begin{center}
      \psfig{file=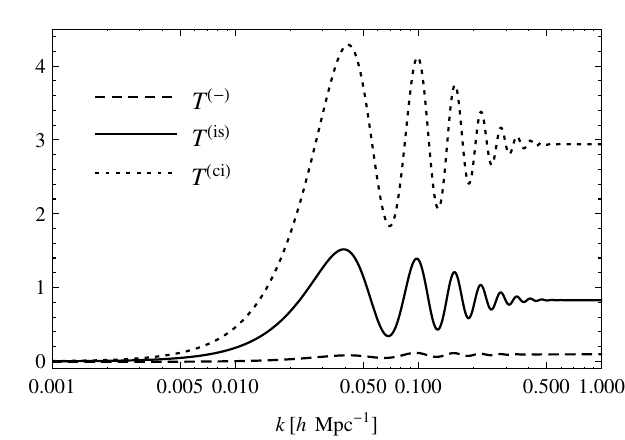,width=8.5cm}
\caption{The transfer functions at $z=900$, normalized to the adiabatic
growing mode. From bottom to top, the adiabatic decaying mode (dashed line), the isodensity  decaying mode (continuous line) and the isodensity constant mode (dotted line).}
\label{ModeTransfer}
\end{center}
\end{figure}

These quantities are shown on Fig.~\ref{ModeTransfer} where we plot the amplitude of the transfer functions $T^{(-)}$, $T^{(\ii)}$ and $T^{(\ci)}$ at initial time, normalized to the amplitude of $T^{(+)}$.
Note that from these results one can compute the r.m.s. of $\Xi^{(\ii)}$ that appeared in the previous section. One finds that
\[
\left\langle {\Xi^{(\ii)}}^2\right\rangle^{1/2} =8.6\cdot 10^{-3} \frac{k}{ h \, {\rm Mpc}^{-1}} \;,
\]
at redshift $z=900$, showing that for our scales of interest, $k\lesssim 1 \, h\, {\rm Mpc}^{-1}$, the effects of the isocurvature modes can only be small. The explicit dependence of the propagators on the isodensity modes is shown in the following.

\subsection{The nonlinear propagators}

In the presence of the decaying isodensity mode, the resummed propagator is no longer proportional to the free field propagator. The effect of the isodensity mode on the resummed propagator is modulated by the matrix $C_{ab} (\eta,\eta_0)$ in eq.~\eqref{correction2}. 
\begin{figure}
\begin{center}
\psfig{file=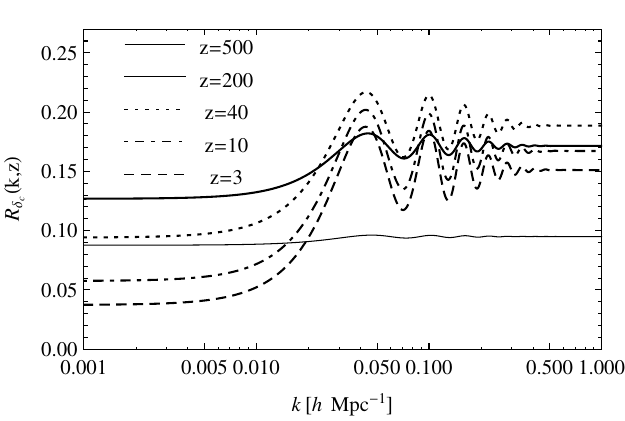,width=8.5cm}
\psfig{file=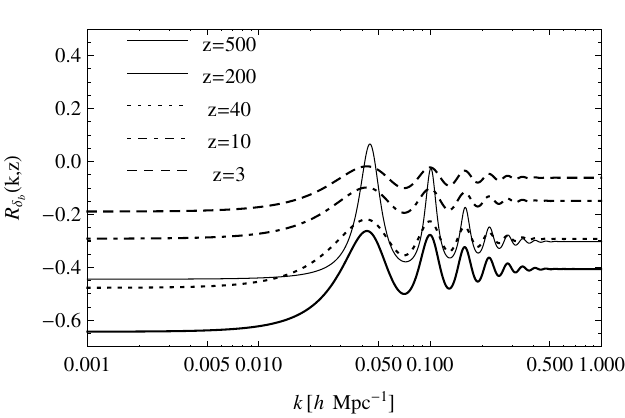,width=8.5cm}
\caption{The quantities $R_{\delta_\cdm}\equiv C_{1a}(z,z_\init=900) T_a(k,z_\init=900) /T_{\delta_\cdm}(k,z) $ (upper pannel) and $R_{\delta_\ba}\equiv C_{3a}(z,z_\init=900) T_a(k,z_\init=900) /T_{\delta_\ba}(k,z) $ (lower pannel) as a function of scale at different redshifts.}
\label{fig:Cab}
\end{center}
\end{figure}
To show this modulation, let us define the quantities
\[
\begin{split}
R_{\delta_\cdm}(k,\eta) & \equiv C_{1a}(\eta,0) T_a(k,0) /T_{\delta_\cdm}(k,\eta) \;, \\ R_{\delta_\ba}(k,\eta) & \equiv C_{3a}(\eta,0) T_a(k,0)  /T_{\delta_\ba}(k,\eta) \;, 
\end{split}
\]
where 
\[
\begin{split}
T_{\delta_\cdm} (k,\eta) & \equiv g_{1a} (\eta,0) T_a(k,0) \;,\\
T_{\delta_\ba}(k,\eta) & \equiv g_{3a} (\eta,0) T_a(k,0)\;.
\end{split}
\]
In Fig.~\ref{fig:Cab} we have plotted these quantities as a function of scale and for different redshifts $z=500, 200, 40, 10, 3$, corresponding to $D_+=1.99, 5.26, 25.71, 95.47, 258.43$. 
\begin{figure}
\begin{center}
            \psfig{file=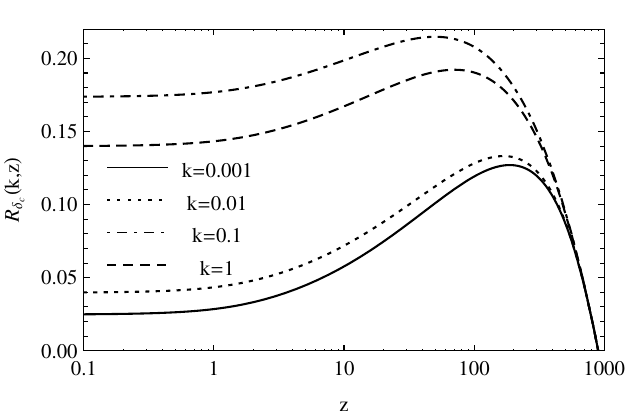,width=8.5cm}
            \psfig{file=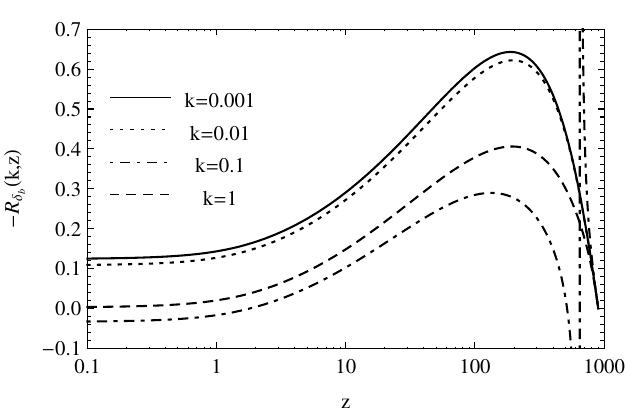,width=8.5cm}
\caption{The quantities $R_{\delta_\cdm}$ (upper pannel) and $R_{\delta_\ba}$ (lower pannel) as a function of redshift at different scales.}
\label{fig:Cab_z}
\end{center}
\end{figure}
In Fig.~\ref{fig:Cab_z} we have plotted $R_{\delta_\cdm}$ and $R_{\delta_\ba}$ as a function of redshift and for different scales $k=0.001,0.01,0.1,1$ in units of $ h \; {\rm Mpc}^{-1}$.
At small redshift (large $\eta$) $R_{\delta_\cdm}$ and $R_{\delta_\ba}$ are dominated by the most growing mode of the matrix $C_{ab}$, i.e.~the first term in eq.~\eqref{correction2} which grows as $e^\eta$, so that they are independent of redshift. At higher redshift the decaying modes in the matrix $C_{ab}$ become important and for $z=900$, corresponding to the initial time $\eta_\init=0$, $R_{\delta_\cdm}$ and $R_{\delta_\ba}$ go to zero.
Note that  $R_{\delta_\ba}$ becomes infinite twice around $z \sim 700$. This is because at early times, right after recombination, the linear baryon density contrast is positive and its decaying isodensity mode dominates over the growing adiabatic mode. Later on it reaches a negative minimum where the growing adiabatic mode starts dominating. Thus $T_{\delta_\ba}(k,\eta)$ crosses zero twice.

The entire effect of the isodensity mode on the propagator is represented by the second term in the square bracket in eq.~\eqref{resummed_prop}, which for $\eta_0=\eta_\init=0$  is 
\[
-k^2 r_\times \sigma_\displ^2 \left(D_+(\eta) -1 \right) C_{ab}(\eta,0)\;.\label{corr_term}
\]
Here the parameter $r_\times$ is the ratio of the isodensity-adiabatic displacement cross-correlation $\sigmax^2$ to the variance of the adiabatic displacement field $\sigma_\displ^2$,
\[
r_\times \equiv \frac{\sigmax^2}{\sigma_\displ^2} =  \frac{{\int} \dd \ln q \;  T^{(\ii)}(q, 0)T^{(+)}(q, 0)q^{n_s-3}}{\int \dd \ln q \;  [T^{(+)}(q, 0)]^2 q^{n_s-3}} \;. 
\]
(In the second equality we have neglected the adiabatic decaying mode.) At $z=900$ this is $r_\times \simeq 0.85$. 
For $k\sigma_\displ D_+ \gg 1 $ the nonlinear propagator goes quickly to zero, due to the exponential damping in eq.~\eqref{resummed_prop}. Thus, the key quantity responsible for suppressing the effect at low redshift is actually the time dependence in eq.~\eqref{corr_term} given by $D_+(z) - 1$. For scales $k < k_\displ \equiv (\sigma_\displ D_+ )^{-1}$, the corrective term \eqref{corr_term} is always found to be 
extremely small. 

\begin{figure}
\begin{center}
            \psfig{file=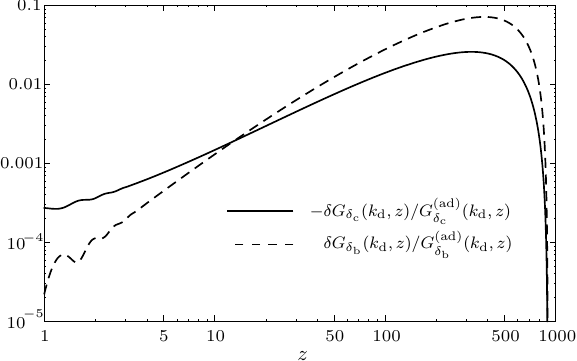,width=8.5cm}
\caption{Effect of the isodensity mode on  the nonlinear propagators normalized to the adiabatic nonlinear propagators,
$\delta G_{\delta_\cdm}/G^{(\rm ad)}_{\delta_\cdm} \equiv
G_{\delta_\cdm}/G^{(\rm ad)}_{\delta_\cdm} -1$
and
$\delta G_{\delta_\ba}/G^{(\rm ad)}_{\delta_\ba}\equiv
G_{\delta_\ba}/G^{(\rm ad)}_{\delta_\ba}  -1$,
where $G^{(\rm ad)}_{\delta_\cdm} $ and $G^{(\rm ad)}_{\delta_\ba} $ are the CDM and baryon nonlinear propagators in absence of isodensity mode,  computed at fixed scale $k_{\displ}$, as a function of redshifts. The oscillations appearing for $z\lesssim3$ are due to the oscillatory behavior of the transfer functions at $k\lesssim0.4$. Note that the effect on the CDM propagator is plotted with the sign changed.}
\label{fig:corr}
\end{center}
\end{figure}
In order to be more quantitative, in Fig.~\ref{fig:corr} we show the effect of the isodensity mode on  the nonlinear propagator by plotting $G_{\delta_\cdm}/G^{(\rm ad)}_{\delta_\cdm} -1$ and $G_{\delta_\ba}/G^{(\rm ad)}_{\delta_\ba}  -1$ as a function of redshift and at fixed scale $k_\displ$, where 
\[
\begin{split}
G_{\delta_\cdm}(k, \eta) &\equiv G_{1a}(k;\eta,0) T_a(k,0)\;, \\
G_{\delta_\ba}(k, \eta) &\equiv G_{3a}(k;\eta,0) T_a(k,0) \;,
\end{split}
\]
and $G^{(\rm ad)}_{\delta_\cdm} (k,z)$ and $G^{(\rm ad)}_{\delta_\ba} (k,z)$ are the same quantities in the adiabatic case -- i.e.~for $\Xi^{(\ii)} =0$. We have used that the value of the variance of the displacement field is $\sigma_\displ \simeq 9.2 \times 10^{-3} h^{-1} $Mpc at $z=900$.
For $z \le 50$ we find that the effect is less that $\sim 1 \%$ and for $z \le 9$ less than $\sim 1$\textperthousand.  Note that the effects are of different signs between CDM and baryons.

Another example where these effects could be significant is when the isodensity modes are set at much lower redshift. This is potentially the case for massive neutrinos. However, massive neutrinos cannot be fully considered as non-relativistic particles during their cosmological history as their behavior is determined by a whole set of extra modes, such as pressure fluctuations and anisotropic stresses. We leave the study of this special case for the future.

\section{Conclusions}

\begin{figure}
\centerline{\epsfig {figure=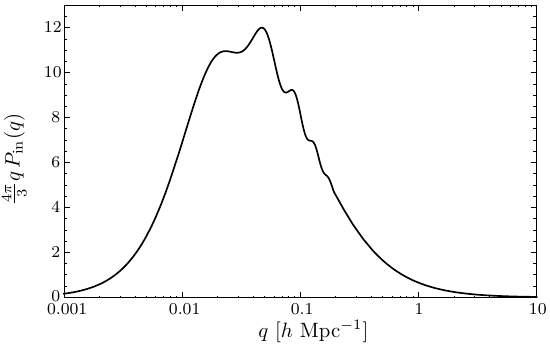,width=8.5cm}}
\caption{Mode contribution per $\log$ interval to the variance of the displacement field from the adiabatic modes, eq.~\eqref{sigmadsq}, at $z=0$.
}
\label{VelDens}
\end{figure}
The {eikonal} approximation provides an efficient  formalism within which exact resummation in the high-$k$ limit can be performed explicitly or numerically. We were able to recover the standard results obtained in \cite{2008PhRvD..78j3521B,2010PhRvD..82h3507B,2006PhRvD..73f3520C,2006PhRvD..73f3519C} concerning the nonlinear 2-point and multi-point propagators describing the gravitational instabilities of a single pressureless fluid. In particular, the propagators are corrected by an exponential cut-off whose scale is fixed by the amplitude of the displacement field along the wave-mode $\vk$. We have shown this irrespective of the growth rate of the displacement field and of whether it follows the linear regime, thus extending the standard results previously quoted. 

Note that this formalism is based on a mode separation between large-scales and small-scales. Indeed, we have assumed that the large-scale modes with momentum $\vq$, collected in the random variable $\Xi$, are much smaller that the small-scale modes $\vk$. In Fig.~\ref{VelDens} we show the contribution from adiabatic modes to the variance of the displacement field, $\sigma_{\displ}$,  per logarithmic interval. As one can see, most of the contribution comes from modes with $q\lesssim 0.1\ h \;{\rm Mpc}^{-1}$ but that of smaller modes, with $q\approx 0.1 \sim 1\ h \;{\rm Mpc}^{-1}$ is not negligible. This suggests that, for $k\approx 0.1\sim 0.3\ h \; {\rm Mpc}^{-1}$, a better description of the damping could be obtained by setting a UV cutoff for $q$ in eq.~\eqref{sigmadsq}.

We have then extended the eikonal approximation to multiple pressureless fluids. In this case one can identify two types of modes: Two adiabatic modes and two isodensity modes per added species. Isodensity modes are responsible for new effects. Indeed, their large-scale flow changes the phase but also (unlike the adiabatic modes) the amplitude of small scales. Thus, the growth of structure and consequently the amplitude of propagators and spectra are affected in a more complex way than in the purely adiabatic case. In this paper we focus our results  on the propagators, leaving the study of power spectra for future work.

In contrast to the single-fluid case, where the effect of large-scale adiabatic modes can be taken into account analytically, for the isodensity modes we have not been able to find an analytic form for the resummed propagator. In this case, one should rely on a numerical or a perturbative approach. The latter is sufficient when one considers the case of CDM-baryon mixing. For this example, we found that the impact of isodensity modes on the propagators is very small at low redshift and for scales of interest for standard PT, i.e.~for $k\lesssim 1 h \; {\rm Mpc}^{-1}$. However, there might be cases where the impact of large-scale modes is more significant, for instance when the scale of interest are  close to the non-linear regime at the time the isodensity modes are set in. This is expected to be the case for massive neutrinos. Although we did not address this case explicitly, we stress that the eikonal method can be used irrespective of the field content of the system. We leave the case of massive neutrinos for further studies.

\

{\bf Acknowledgements:} We thank the participants of the PT{chat} workshop at IPhT in Saclay for interesting discussions. FV wishes to thank Antonio Riotto and Ravi Sheth for fruitful conversations.

\appendix
\section{Computation of moments}
\label{moments}
We want to compute the value of
\[
g_{n}=\langle \Xi^{n} \exp(\mD)\rangle \;,
\]
where $\Xi$ and $\mD$ are two random variables whose statistical properties are entirely characterized
by their joint cumulants,
\[
x_{p,q}=\langle \Xi^{p}\mD^{q}\rangle_{c} \;.
\]
It is convenient to introduce the auxiliary function $\exp(\mD+\lambda \Xi)$ and to notice that
\[
g_{n}=\frac{\dd^{n}}{\dd \lambda^{n}}\left.\langle \exp(\mD+\lambda \Xi)\rangle\right\vert_{\lambda=0} \;.
\]

The ensemble average that appears in this expression can be written in terms of the cumulants generating function
of $\mD+\lambda \Xi$ as
\[
g_{n}=\frac{\dd^{n}}{\dd \lambda^{n}}
\left.\exp\left(
\sum_{p=0}^{\infty}\sum_{q=0}^{p}\frac{1}{q!}\frac{1}{(p-q)!}x_{q,p-q}\lambda^{q}
\right)\right\vert_{\lambda=0} \;,
\]
which can be re-written as
\[
g_{n}=\exp\left(\sum_{p=0}^{\infty} \frac{x_{0,p}}{p!}\right)
\frac{\dd^{n}}{\dd \lambda^{n}}
\left.\exp\left(\sum_{q=1}^{n} \frac{X_{q}}{q!}\lambda^{q}\right)\right\vert_{\lambda=0}\;,
\]
where
\[
X_{q}=\sum_{p=q}^{\infty}\frac{1}{(p-q)!}x_{q,p-q} \;.
\]
$g_n$ can then be formally expressed in terms of $X_{q}$ as
\begin{align}
g_{0}&=\exp\left(\sum_{p=0}^{\infty} \frac{x_{0,p}}{p!}\right) \;, \\
g_{1}&=X_{1}\ g_{0}\; ,\\
g_{2}&=\left(X_{1}^{2}+X_{2}\right)\ g_{0}\; ,\\
g_{3}&=\left(X_{1}^{3}+3 X_{1}X_{2}+X_{3}\right)\ g_{0}\; ,\\
&\dots \;. \nonumber
\end{align}
Note that for a unit $g_{0}$ the relation between $X_{q}$ and $g_{n}$ is exactly the one relating cumulants of order
$q$ with moments of order $n$. Thus, in general this relation is obtained by the Arbogast-Fa\`a di Bruno formulae.

These relations greatly simplify in the case of Gaussian initial conditions, assuming -- without  loss of generality -- that $x_{0,0}=x_{1,0}=x_{0,1}=0$. Indeed, in this case only $X_{1}$ and $X_{2}$ are non-zero,
\[
X_{1}=x_{1,1}\;, \quad X_{2}=x_{2,0} \;,
\]
so that
\begin{align}
g_{0}&=\exp\left(\frac{x_{0,2}}{2!}\right)\; ,\\
g_{1}&=x_{1,1}\ g_{0}\; ,\\
g_{2}&=\left(x_{1,1}^{2}+x_{2,0}\right)\ g_{0}\; ,\\
g_{3}&=\left(x_{1,1}^{3}+3 x_{1,1}x_{2,0}\right)\ g_{0}\; ,\\
&\dots \;.\nonumber
\end{align}

\section{Higher-order time  dependence}
\label{HOTimeDependence}

It is possible to compute the nonlinear propagator at higher orders in $\Xi^{(\ii)}$ using the expansion
\begin{widetext}
\[
\xi_{af} (\vk, \eta,\eta_0) = 
g_{af}(\eta,\eta_0) \exp \left( \int^{\eta}_{\eta_0}\dd\eta' \Xi^{({\rm ad})}(\vk,\eta') \right)+ {\xi}^{(1)}_{af} (\vk,\eta,\eta_0) + {\xi}^{(2)}_{af}(\vk,\eta,\eta_0) + \ldots 
\;, \label{xi_ho}
\]
where
\[
\begin{split}
{\xi}^{(1)}_{af} & \equiv \exp \left( \int^{\eta}_{\eta_0}\dd\eta' \Xi^{({\rm ad})}(\vk,\eta') \right) \int_{\eta_0}^\eta \dd \eta' g_{ab}(\eta,\eta') \Xi_{bc}^{(\ii)} (\vk,\eta') g_{cf} ( \eta',\eta_0) \;, \\
{\xi}^{(2)}_{af} & \equiv \exp \left( \int^{\eta}_{\eta_0}\dd\eta' \Xi^{({\rm ad})}(\vk,\eta') \right) \int_{\eta_0}^\eta \dd \eta' \int_{\eta_0}^{\eta'} \dd \eta'' g_{ab}(\eta,\eta') \Xi_{bc}^{(\ii)} (\vk,\eta') g_{cd} ( \eta',\eta'') \Xi_{de}^{(\ii)} (\vk,\eta'') g_{ef} ( \eta'',\eta_0)\;, \\
& \ldots  \;.
\label{integrals}
\end{split}
\]
\end{widetext}
In particular, here we derive an explicit expression for the most growing solution to the, in terms of a recurrence formula.

The key point is to show that the fastest growing mode of the amplitude of the corrections to the resummed propagator goes as the adiabatic growing mode, i.e.~$\propto e^{\eta-\eta_0}$, to all orders in $\Xi^{(\ii)}$. Let us consider the amplitude of the $n^\textrm{th}$-order correction. From eq.~\eqref{integrals} and using that $g_{ab}^{(\lambda)} (\eta,\eta_0) \propto e^{\lambda (\eta - \eta_0)}$ and $\Xi_{ab}^{(\ii)}(\eta) \propto e^{-\frac12 \eta}$  ($\eta_\init=0$), its time dependence is proportional to  $n$ nested integrals, 
given by
\begin{widetext}
\[
{\cal I}^{(n)} \equiv
\int_{\eta_0}^{\eta}
      \dd\eta_1\ e^{\lambda_0(\eta -\eta_1)} e^{-\frac12 \eta_1}
\int_{\eta_0}^{\eta_1}
      \dd\eta_2\ e^{\lambda_1(\eta_1-\eta_2)}
      e^{-\frac12 \eta_2 }
\cdots
\int_{\eta_0}^{\eta_{n-1}}
      \dd\eta_n\ e^{\lambda_{n-1}(\eta_{n-1}-\eta_n)}
      e^{-\frac12 \eta_n}
e^{\lambda_{n}(\eta_n-\eta_0)}\;, \label{integrals_2}
\]
\end{widetext}
where each $\lambda_i$ can take the values $\{1,0,-1/2,-3/2\}$. As we are only interested in the fastest growing mode, we take $\lambda_0=1$.  
      
We define
\[
\alpha_{ij}\equiv\lambda_j-\lambda_i-\frac{j-i}2 \;,
\] 
satisfying
$\alpha_{ij}+\alpha_{jk}=\alpha_{ik}$. It is important to note that $\alpha_{0i}<0$ for $i\neq0$. Then, we can rewrite the integrals above as
\[
C
\int_{\eta_0}^{\eta}\dd\eta_1\ e^{\alpha_{01} \eta_1}
\int_{\eta_0}^{\eta_1}\dd\eta_2\ e^{\alpha_{12} \eta_2}
\int_{\eta_0}^{\eta_{n-1}}\dd\eta_n\ e^{\alpha_{n-1,n} \eta_n}
\]
with $C \equiv e^{\lambda_0\eta}e^{-\lambda_n \eta_0} $\;.
By performing an integration by parts on $\eta_1$, we find the boundary term
\[
C\left[
\frac{e^{\alpha_{01}\eta_1}}{\alpha_{01}}
\int_{\eta_0}^{\eta_1}\dd\eta_2\
e^{\alpha_{12}\eta_2}\cdots\int_{\eta_0}^{\eta_{n-1}}\dd \eta_n\
e^{\alpha_{n-1,n} \eta_n}
\right]_{\eta_1=\eta_0}^{\eta_1=\eta} \;,
\]
and the remaining integrals
\[
-C\int_{\eta_0}^\eta\dd\eta_1\ \frac{e^{\alpha_{01}\eta_1}}{\alpha_{01}}
e^{\alpha_{12}\eta_1}\cdots\int_{\eta_0}^{\eta_{n-1}}\dd
\eta_n\ e^{\alpha_{n-1,n} \eta_n}\;.
\]
The lower bound for the boundary term obviously vanishes. Therefore, the boundary term contains an $e^{\alpha_{01}\eta}$ factor making it subleading with respect to the remaining integrals, which for large $\eta$ become constant. To see this, we perform $(n-2)$ more integration by parts, every time dropping the boundary term for the very same reason, until we are left with
\[
(-1)^{n-1}
\frac{C}{\alpha_{01}\alpha_{02}\cdots\alpha_{0,n-1}}
\int_{\eta_0}^{\eta}\dd\eta_n\ e^{\alpha_{0n}\eta_n} \;.
\]
The leading term of this integral is independent of $\eta$,
\[
(-1)^n\prod_{i=1}^n
\frac{C}{\alpha_{0i}} e^{\alpha_{0n}\eta_0}\;,
\]
and by replacing $C$ by its definition, we have
\[\label{I(n)sol}
\mathcal
{\cal I}^{(n)} =
\prod_{i=1}^n \dfrac{1}{ \alpha_{i0}}
e^{- \frac{n}2 \eta_0}\,e^{\eta-\eta_0}.
\]

Now that we have studied the time dependence, we can reintroduce the time independent matrices $h_{ab} g_{bc}^{(\lambda_i)}$ in \eqref{integrals} and sum over $\lambda_i$ in each of the integrals of eq.~\eqref{integrals_2}.
Let us define the time independent matrices
\[
\begin{split}
A_{ab}^{(0)} & \equiv  g_{ab}^{(1)} \;,\\
A_{ab}^{(i)} & \equiv  \sum_{\lambda=1,0,-1/2,-3/2} \frac{  h_{ac} \, g_{cb}^{(\lambda)} }{1+i/2 - \lambda}\;, \quad i \ge 1\;, \label{rec_formula}
\end{split}
\]
where $h_{ab}$ is defined in eq.~\eqref{iso_prop}, $g_{ab}^{(\lambda)} \equiv g_{ab}^{(\lambda)}(\eta_0,\eta_0)$ are the time independent projectors on the right-hand side of eq.~\eqref{linear_g}, and the sum runs over $\lambda=1,0,-1/2,-3/2$ corresponding to the 4 linear modes $+$, $\ci$, $\ii$ and $-$. The denominator $(1+i/2-\lambda)$ is exactly the $\alpha_{i0}$ of eq.~\eqref{I(n)sol}.
It is easy to verify that the most growing solution for $\xi_{ab}^{(n)}$ in  eq.~\eqref{xi_ho} is then given by
\[
\begin{split}
{\xi}^{(n)}_{ab}(\vk,\eta,\eta_0) =& \left[ \Xi^{(\ii)} (\vk,0) e^{-\eta_0/2} \right]^n \Big[ {\bf A}^{(0)}\prod_{i=1}^{n} {\bf A}^{(i)} \Big]_{ab}  \\
&\times e^{\eta-\eta_0} \exp \left( \int^{\eta}_{\eta_0}\dd\eta' \Xi^{({\rm ad})}(\vk,\eta') \right)\;. 
\end{split}
\]
This equation formally generalizes the expression for the most growing mode contained in eqs.~\eqref{correction1} and \eqref{correction2}  to any order. Note that the time dependence is the same as the \textsl{linear} growing mode, i.e.~$\propto e^{\eta}$. Ensemble averages of these quantities can 
be taken using the equations given in appendix \ref{moments}.

\section{Treating super-horizon scales}
\label{SH}

Let us consider a linearly perturbed FLRW metric in longitudinal gauge with only scalar perturbations. In the absence of anisotropic stress the two metric potentials are identical and the metric simply reads
\[
\dd s^2 = a^2(\tau)\left[ - (1+2 \phi_\lon) \dd \tau^2 + (1-2 \phi_\lon) \dd \xv^2 \right]\;,
\]
where $\tau$ is the conformal time defined by $\dd \tau = \dd t /a(t)$.

Combining the $00$ and the $0i$ components of Einstein's equations gives, in Fourier space, \cite{Ma1995}
\[
k^2 \phi_\lon= - \frac32 \HH^2 \left(\delta_\lon + 3 \HH^2 \theta_\lon /k^2 \right)\;, \label{Poisson_GR}
\]
where $\delta_\lon$ and $\theta_\lon$ are the total density contrast and dimensionless velocity divergence \footnote{Note that contrary to the notation of \cite{Ma1995}, here $\theta$ denotes the \textsl{dimensionless} velocity divergence so that $\theta_{\rm our} \equiv \theta_{\rm MB}/\HH$.}, respectively, in longitudinal gauge and ${\cal H} \equiv \dd \ln a/\dd \tau$ is the conformal Hubble rate.
Moreover, in this gauge, the continuity and Euler equations for a single-fluid read, at linear order, \cite{Ma1995}
\begin{align}
\delta_\lon' &= - \HH \theta_\lon+ 3 \phi_\lon ' \;, \label{continuity_notes}\\
\theta_\lon' &= -( \HH'/\HH + \HH) \theta_\lon + k^2 \phi_\lon/\HH\;, \label{euler_notes}
\end{align}
where a prime denotes the derivative with respect to conformal time.

One can check that, at the linear level, the Newtonian Euler equation \eqref{Euler_tot} is the same as its relativistic version, eq.~\eqref{euler_notes}, while the continuity equations \eqref{continuity} and \eqref{continuity_notes} differ by the term $3\phi_\lon '$. Indeed, for instance, using that $a\propto \tau^2$ in matter dominance, one can check that the solutions to the above equations are
\[
\phi_\lon (k, \tau)= \phi_+(k) +  (k \tau)^{-5} \phi_- (k)\;,
\]
and
\begin{align}
\delta_\lon = & - \left(2  + \frac{1}{6} (k \tau)^2 \right) \phi_+ \nonumber \\
 &-   \left( \frac16 (k \tau)^{-3}  - 3 (k \tau)^{-5} \right)\phi_- \;, \\
\theta_\lon  = & \ \frac16 (k \tau)^2 \; \phi_+  - \frac14  (k  \tau)^{-3} \; \phi_- \;. \label{theta_long}
\end{align}
Thus, eq.~\eqref{theta_long}  correctly describes the growing and decaying solutions of $\theta$ in the Newtonian limit, eq.~\eqref{theta+-}, with $\theta_+ \propto a $ and $\theta_- \propto a^{-3/2} $, even on super-Hubble scales, while for $\delta_\lon$ we recover the Newtonian case, eq.~\eqref{delta+-}, only in the limit $k \tau \gg 1$.

One can define a quantity which obeys the Newtonian continuity equation even on super-Hubble scales. The comoving energy density perturbation, defined as
\[
\begin{split}
\delta_{\rm com} & \equiv \delta_\lon + 3 \HH^2 \theta_\lon /k^2 \\ &= -\frac16 \left( (k \tau)^2 \phi_+ +  (k \tau)^{-3} \phi_- \right)\;,
\end{split}
\]
does this job. Indeed, replacing $\phi_\lon'$ using the $0i$ components of Einstein's equation \cite{Ma1995}, and using eq.~\eqref{euler_notes}, eq.~\eqref{continuity_notes} reads
\[
\delta_{\rm com}' = - \HH \theta_\lon\;,
\]
thus reproducing the linear part of the continuity equations in the Newtonian limit, eq.~\eqref{continuity}.
Moreover, in terms of this variable eq.~\eqref{Poisson_GR} becomes a Poisson-like equation,
\[
k^2 \phi_\lon= - \frac32 \HH^2 \delta_{\rm com}\;,
\]
reproducing eq.~\eqref{poisson}.

We conclude that at linear level the Newtonian equations \eqref{continuity}, \eqref{Euler_tot} and \eqref{poisson} describe the relativistic dynamics once we interpret the Newtonian potential $ \phi$ as the metric potentials in longitudinal gauge $\phi_\lon$, the Newtonian density contrast $\delta$ as the \textsl{comoving} energy density perturbation $\delta_{\rm com}$ and $\theta$ as the dimensionless velocity divergence in longitudinal gauge, $\theta_\lon$.

The case of many fluids is not very different. In this case one can show that for each species $\alpha$ the relativistic perturbation variable which satisfies the Newtonian continuity equation is the energy density perturbation comoving to the \textsl{total} fluid,
\[
\delta_{\alpha, {\rm com}} \equiv \delta_{\alpha,\lon} + 3 \HH^2 \theta_\lon/k^2\;,
\]
where $\theta_\lon$ is the total  dimensionless velocity divergence in longitudinal gauge,
\[
\theta_\lon \equiv \sum_\alpha f_\alpha \theta_{\alpha,\lon}\;.
\]
Using this variable, eq.~\eqref{Poisson_GR} becomes a Poisson equation while the velocity divergence in longitudinal gauge for each species $\alpha$, $\theta_{\alpha,\lon}$, naturally satisfies the Newtonian Euler equation.

Let us connect these variables with those of CAMB in the case of a mixture of cold dark matter and baryons. CAMB uses synchronous gauge comoving with cold dark matter \cite{Lewis:1999bs}. Using the gauge transformation between synchronous and  longitudinal gauge \cite{Ma1995} we have that 
\[
\begin{split}
\delta_{\alpha, \rm CAMB} & = \delta_{\alpha, \lon} + 3 \HH \sigma\;,\\
\theta_{\alpha, \rm CAMB} &= \theta_{\alpha, \lon} - \sigma k^2/\HH \;,
\end{split}
\]
where $\sigma$ is the shear. This is related to the usual synchronous metric perturbation variables $h_{\rm syn}$ and $\eta_{\rm syn}$, respectively the trace and traceless part of the spatial metric, by $\sigma = ( h_{\rm syn}' + 6 _{\rm syn}')/2k^2$. Using these equations, since in CAMB the dark matter velocity vanishes, $\theta_{\cdm, \rm CAMB} = 0$, we have
\[
\begin{split}
\theta_{\cdm, \lon} &= \sigma k^2/\HH\;, \\
\theta_{\ba, \lon} &= \theta_{\ba, \rm CAMB}+ \sigma k^2/\HH\;, \\
\delta_{\alpha,{\rm com}}  & = \delta_{\alpha, \rm CAMB} + 3 \HH^2 f_\ba  \theta_{\ba,\rm CAMB} /k^2\;.
\end{split}
\]

\

\bibliography{Eikonal}

\end{document}